\documentclass{aa}  

\usepackage{graphicx}
\usepackage{txfonts}
\usepackage{xspace}
\usepackage{color}
\usepackage[dvipsnames]{xcolor}
\usepackage{multirow}
\usepackage{hyperref}
\usepackage{soul}
\usepackage{ulem}
\usepackage{pdflscape}
\usepackage{afterpage}

\newcommand{\drv}[0]{$\Delta RV_{\rm max}$\xspace}
\newcommand{\qcf}[0]{$q_{\rm set}$\xspace}

\newcommand{\qc}[0]{$q_{\rm crit}$\xspace}

\begin{document}

   \title{Calibration of Binary Population Synthesis Models Using White Dwarf Binaries from APOGEE, GALEX and Gaia}

     \titlerunning{Calibration of Binary Population Synthesis Models with WD Binaries from AGGC}

   \author{A. C. Rubio\inst{1,2}, K. Breivik\inst{3},
            C. Badenes\inst{4},
          K. El-Badry\inst{5},
          B. Anguiano\inst{6,7},
          E. Linck\inst{8},
          S. Majewski\inst{7},
          K. Stassun\inst{9,10}}
    
   \institute{Max-Planck-Institut für Astrophysik, Karl-Schwarzschild-Str. 1, 85748 Garching b.\ M\"unchen, Germany
   \and 
   Instituto de Astronomia, Geof{\' i}sica e Ci{\^e}ncias Atmosf{\'e}ricas, Universidade de S{\~ a}o Paulo, Rua do Mat{\~ a}o 1226, Cidade Universit{\' a}ria, 05508-900 S{\~a}o Paulo, SP, Brazil
        \and
        Department of Physics, McWilliams Center for Cosmology and Astrophysics, Carnegie Mellon University, 5000 Forbes Avenue, Pittsburgh, PA 15213, USA
        \and
        Department of Physics and Astronomy and Pittsburgh Particle Physics, Astrophysics and Cosmology Center (PITT PACC), University of Pittsburgh, 3941 O’Hara Street, Pittsburgh, PA 15260, USA\and
        Department of Astronomy, California Institute of Technology, 1200 E. California Blvd., Pasadena, CA 91125, USA\and
        Centro de Estudios de Física del Cosmos de Aragón (CEFCA), Plaza San Juan 1, 44001, Teruel, Spain\and
        Department of Astronomy, University of Virginia, Charlottesville, VA, 22904, USA\and
        University of Wisconsin-Madison, 2535 Sterling Hall 475 N. Charter Street Madison, WI 53706, USA\and
        Department of Physics and Astronomy, Vanderbilt University, Nashville, TN 37235, USA\and
        Department of Physics, Fisk University, Nashville, TN 37208, USA
        \\
        \email{amanda.rubio@usp.br - rubio@mpa-garching.mpg.de}
          }   

   \date{Received xx; accepted xx}

 
  \abstract
  {The effectiveness and stability of mass transfer in a binary system are crucial in determining the final product of its evolution. 
  Rapid binary population synthesis codes simplify the complex physics of mass transfer and common-envelope evolution by adopting parameterized prescriptions for the stability of mass transfer, accretion efficiency in stable mass transfer, and the efficiency of common-envelope ejection. 
Our goal is to calibrate these uncertain parameters by comparing binary population synthesis models with observational data. Binary systems composed of a white dwarf and main sequence star are an ideal population to study the effects of binary interaction, as they can be formed through stable or unstable mass transfer, or without any interaction. These different evolutionary paths affect the orbital period and masses of the present-day population. The APOGEE-GALEX-Gaia catalog (AGGC) provides a homogeneous sample of over 500 systems with well measured radial velocities that can be used as a comparison baseline for population synthesis simulations of white dwarf -- main sequence binaries. 
We compare the distribution of observed maximum radial velocity variation (\drv) in the AGGC to binary population models simulated with COSMIC, a publicly developed binary population synthesis code. Within these synthetic populations, we vary the mass transfer and common-envelope ejection efficiency, and the criteria for mass transfer stability at the first ascent, asymptotic, and thermally pulsing giant branch phases. We also compare our models to systems with orbital solutions and estimated stellar masses.
The \drv comparison shows clear preference for models with a higher fraction of stars undergoing stable mass transfer during the first ascent giant branch phase, and for highly effective envelope ejection during common-envelope phases. For the few systems with estimated WD masses, comparison to models shows a slight preference for non-conservative stable mass transfer.
In COSMIC and similar codes, the envelope ejection efficiency and the envelope binding energy are degenerate parameters. Therefore, our result of high ejection efficiency may indicate that either additional sources of energy (such as recombination energy from the expansion of the envelope) are required to eject the envelope, or that its binding energy is lower than traditionally assumed. Future comparisons to population synthesis simulations of WD binaries can be drawn for other datasets as they become available, such as upcoming Gaia data releases and the future LISA mission, and for binary systems in other evolutionary stages. }

\authorrunning{A. C. Rubio, K. Breivik, C. Badenes}
   \keywords{binaries: general - binaries: close - binaries: spectroscopic - white dwarfs
               }

   \maketitle
%

\section{Introduction}\label{sect:intro}

   White dwarfs (WDs) are the final stage of evolution for the majority of stars. While they are a straightforward end-point for stars with masses below the threshold for central oxygen ignition ($\sim8 \, M_{\odot}$), WDs can also be formed via binary evolution and even be the result of mergers \citep{Althaus2010}. A binary system containing a WD can be formed through many different channels. Two stars can be born in a binary system but be separated enough to never interact. As they evolve as single stars, the result can be a WD and a Main Sequence (MS), giant, or another WD, depending mostly on their initial mass ratio and the age of the system. If the two stars are born close enough to interact, a WD can also be formed after a period of mass transfer, which can be stable or unstable. In stable mass transfer, the WD progenitor increases its radius and fills its Roche lobe. Depending on the stage of evolution of the two stars and their separation, this mass transfer can become unstable, leading to common envelope evolution (CEE), also resulting in a WD binary with closer orbital separations when compared to stable mass transfer \citep{Ivanova2013,toonen2017}.

   Regardless of the evolutionary path taken by each WD binary, the physical processes that mediate their formation should leave an imprint on the properties of the final population, such as their period and mass distributions. Thus, studies of large samples of WD binaries can provide information on binary stellar evolution, in particular on the still poorly understood CEE \citep[e.g.,][]{zorotovic2010,Zorotovic2011,deMarco2011,Postnov2014,scherbak2023}. 
   The two vital ingredients for this type of study are data and models that can be readily compared to each other, so that the stability and efficiency of mass transfer can be constrained. For instance, \citet{zorotovic2010}, uses a sample of WD+MS binaries from SDSS to estimate the efficiency of envelope ejection after CEE, $\alpha$, finding values in the range of $0.2 - 0.3$ to represent the final parameters of most of their sample. The recent work of \citet{scherbak2023} models the CEE of 10 WD binary systems using stellar evolution code MESA, finding a similar range for $\alpha$, $0.2 - 0.4$. 
   
   To understand the behavior of an entire population, however, the sample size of the data and the models must be large enough to be statistically relevant. The great multi-epoch surveys of the past decade have finally provided the community with larger, less-biased datasets; combining such surveys allows for assembly of curated catalogs of populations such as WD binaries. The recent APOGEE-Galex-Gaia catalog (AGGC) \citep{anguiano2022} is one such effort to minimize biases, providing a sample of WD binaries selected with homogeneous, and relatively simple, criteria. On the modeling side, rapid binary population synthesis (BPS) codes \citep{hurley2002} can simulate large populations in a time- and resource-efficient fashion when compared to detailed stellar evolution modeling. These BPS codes use parametrized prescriptions for binary interactions at each step of the evolution of the system, following it from the zero-age main sequence to the final product. Thus, with observational and simulated populations in hand, the theoretical predictions can be compared to data, informing these widely used prescriptions and providing a better understanding of binary evolution.

   In this work, we calibrate the mass transfer parameters in the BPS code COSMIC \citep{breivik2020} using a sample of WD+MS binaries from AGGC \citep{anguiano2022}. In our COSMIC models, we explore different assumptions for the criteria for the onset of unstable mass transfer, limits on the accretion between stars during stable Roche-lobe overflow (RLO), and the efficiency of envelope ejection in CEE. These models are compared to the WD+MS systems in the AGGC by their radial velocity shifts, orbital parameters and stellar mass. Our results provide new estimates on the mass transfer parameters for WD+MS binaries, and are applicable also to other BPS codes that rely on the formalism of \citet{hurley2002}.

The rest of the paper is structured as follows: Section~\ref{sect:data} describes the data contained in the AGGC while Section~\ref{sect:cosmic} describes our BPS methodology. We discuss our results in Section~\ref{sect:results} and compare them to previous work in Section~\ref{sec:discussion}. We finish with our conclusions in Section~\ref{sec:conclusion}.

\section{APOGEE-Gaia-Galex catalog (AGGC)}\label{sect:data}

The catalog was compiled by \citep{anguiano2022} using near-infrared data from the 17th Data Release of the Apache Point Observatory Galactic Evolution Experiment (APOGEE data release 17) \citep{majewski2017, abdurro'uf2022}, ultraviolet (UV) data from GALEX \citep{bianchi2017}, and optical data from the Gaia mission \citep{lindegren2021}. With this combination, \citet{anguiano2022} was able to identify WDs in systems with primary stars in a broad range of luminosities, from low mass MS stars to giants. The APOGEE-Gaia-Galex catalog (AGGC) was compiled by cross-matching all the targets in APOGEE for which the APOGEE Stellar Parameters and Chemical Abundances Pipeline yielded valid atmospheric measurements with the GALEX \citep{bianchi2017} and WISE \citep{wright2010} databases. A full spectral energy distribution (SED) was derived for all the entries that had complete photometric information: two GALEX bands in the UV, four WISE bands in the mid infrared, plus the three Gaia bands in the optical and the three 2MASS bands in the near-infrared that are already part of the APOGEE database. Binary WD candidates were drawn from this sample by requiring an APOGEE $T_{\rm eff}<6000$ K and a GALEX far-UV - near-UV measure of (FUV–NUV)$_{\circ}$ < 5. 

\begin{figure*}
    \centering
    \includegraphics[scale=0.75]{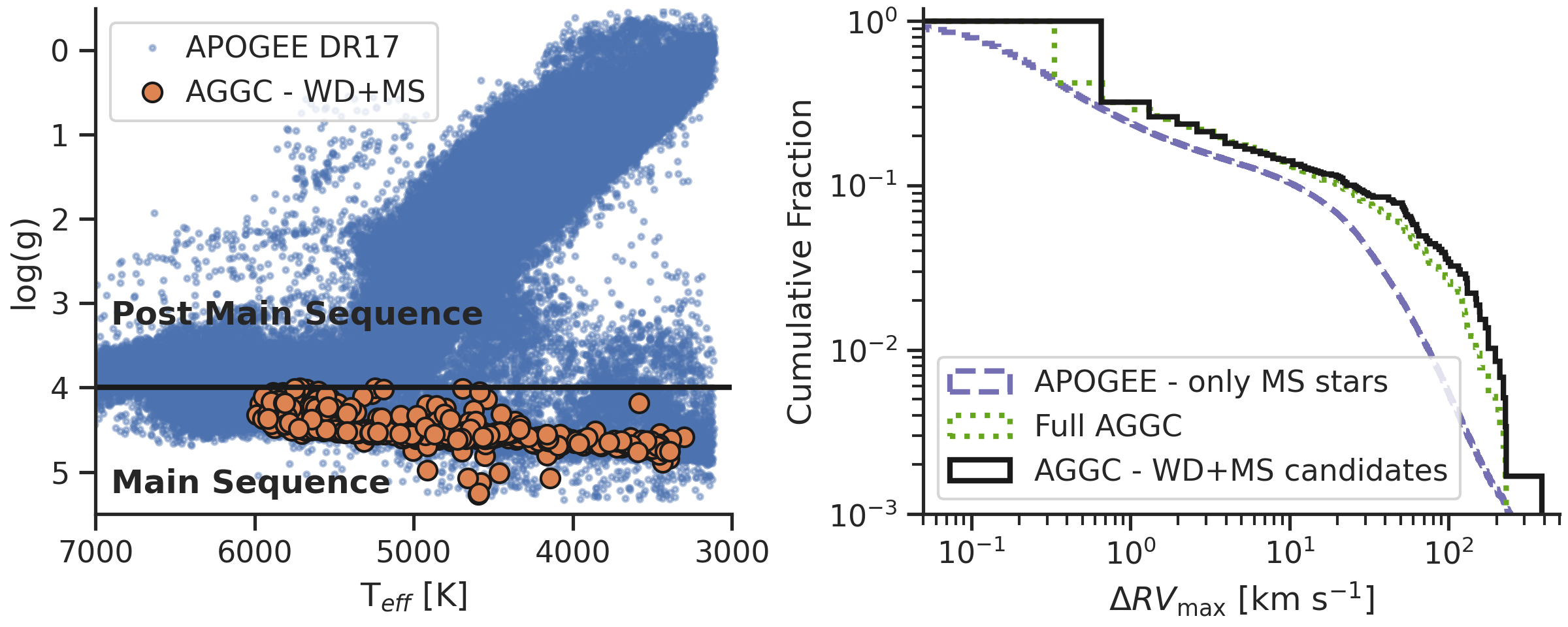}
    \caption{Overview of the white dwarf binaries in the APOGEE-Gaia-Galex catalog (AGGC). The left panel shows the full APOGEE dataset in blue and the companions of WDs in orange. The right panel shows the \drv distribution for different cuts in the data: MS from the full APOGEE in blue, all WD binaries in the AGGC in green, and WD+MS binaries in black. The full APOGEE dataset contains 455796 targets; the MS stars in that sample number 151266. The full AGGC has 1157 candidate WD binaries, while the WD+MS systems number 588. }
    \label{fig:data}
\end{figure*}

The contamination of sample the of WD binary candidates was reduced by removing known intruders (pulsating stars and variable stars) and performing two-component fits to the broadband SEDs, with a `cold' component fixed to a Kurucz model with the APOGEE stellar parameters ($T_{\rm eff}$, [M/H], and $\log g$), and a `hot' component drawn from the Koester grid of atmospheric WD models \citep{koester2010}. The final sample of WD binary candidates in the AGGC was drawn by requiring that the equivalent radius of the WD, $R_{\rm WD}$, inferred from the SED fit be < 25 $R_{\oplus}$. 
Below the limit of 5 $R_\odot$ for the radius of the non-WD companion, the maximum $R_{\rm WD}$ decreases following a slope of 0.7 + 6$R$ of the non-WD stars (see the red lines in Figure~10 of \citealt{anguiano2022}). In the upper left panel of Figure~\ref{fig:data} we show the {observational Hertzsprung-Russell} diagram for the entire APOGEE sample in blue, with the {588} WD+MS binaries in the AGGC highlighted in orange. The black line defines {our applied surface gravity cutoff for the MS companions} ($\log\,g < 4$). 
Figure~\ref{fig:data} provides an overview of the observational data we use in this study, the candidate WD+MS binaries contained in the catalog.

One key advantage of the AGGC is that the APOGEE data provides multi-epoch, precise radial velocities ($\Delta RV \approx$ 100 m s$^{-1}$) for all objects in the sample. In most cases, the number of epochs available is not sufficient to derive a full orbital solution, but the maximum value of the RV variation (maximum minus minimum RV for each target), \drv, is a robust proxy for the orbital period --- see \citep{badenes2018}, \citep{mazzola2020}, and \citep{anguiano2022} for discussions. These \drv measurements are an excellent figure of merit to single out short period binaries, as even accounting for random inclinations on the plane of the sky the highest \drv values will always be associated with the shortest orbital periods, and vice versa. Thus, \drv is a noisy, but unbiased, estimator for the orbital period. In the right panel of Figure~\ref{fig:data}, we compare \drv values for all MS stars in APOGEE (green dotted lines), the AGGC without a MS companion cut (blue dotted lines), and the WD-MS population in the AGGC (black solid lines). 
There is a stark difference in the shape of the cumulative distribution of \drv between the AGGC WD+MS binaries and the full APOGEE dataset. The fact that the AGGC WD+MS binaries have a \drv distribution skewed toward higher values indicates that the binaries in this sample have shorter periods than those of the APOGEE targets that show no evidence for a WD companion.

\begin{table*}[!ht]
    \centering
    \begin{tabular}{l|ccccccr}
    \hline
        APOGEE ID & Parallax [mas] & Period [days] & \drv [km/s] & $M_{\rm MS} \, [M_{\odot}]$  & min $M_{\rm WD} [M_{\odot}]$ & Solution \\
        \hline
        \hline
        2M00471103+1738040 & 11.956 $\pm$ 0.022 &  1.245 $\pm$ 1.23E-05 & 62.914 & 0.522 $\pm$ 0.084 & 0.266 $\pm$ 0.025 & SB1 \\ 
        2M01592351-0045463 & 8.155 $\pm$ 0.026 &  571.951 $\pm$ 0.867 & 22.235 & 0.858 $\pm$ 0.110 & 0.532 $\pm$ 0.055 & Astrometry \\ 
        2M07515777+1807352 & 7.280 $\pm$ 0.021 & 10.298 $\pm$ 6.60E-04 & 58.202 & 0.670 $\pm$ 0.082 & 1.067 $\pm$ 0.058 & SB1 \\ 
        2M08110666+2111127* & 7.089 $\pm$ 0.023 & 202.551 $\pm$ 0.817 & 3.799 & 0.550 $\pm$ 0.021 & 0.177 $\pm$ 0.024 & Astrometry \\
        2M08180299+4412265 & 6.232 $\pm$ 0.026 &  1340.662 $\pm$ 67.723 & 0.133 & 0.596 $\pm$ 0.100 & 0.550 $\pm$ 0.043 & Astrometry \\ 
        2M08410922+2542362 & 3.338 $\pm$ 0.027 & 464.087 $\pm$ 1.731 & 0.086 & 1.000 $\pm$ 0.124 & 0.506 $\pm$ 0.050 & Astrometry \\ 
        2M11463394+0055104 & 4.854 $\pm$ 0.083 & 0.409 $\pm$ 1.14E-05 & 373.00 $\pm$ 2.07 & 0.440 $\pm$ 0.028 & 0.717 $\pm$ 0.014 & RV curve \\
        2M14544500+4626456 & 9.054 $\pm$ 0.028 & 15.096 $\pm$ 6.98E-05 & 99.53 $\pm$ 0.04 & 0.488 $\pm$ 0.031 & 0.693 $\pm$ 0.001 & RV curve \\
        2M15460673+2655088 & 3.167 $\pm$ 0.017 &  1105.316 $\pm$ 34.542 & 0.359 & 0.809 $\pm$ 0.105 & 0.580 $\pm$ 0.125 & Astrometry \\ 
        2M16105316+4130067* & 8.316 $\pm$ 0.012 &  204.018 $\pm$ 0.200 & 5.482 & 0.883 $\pm$ 0.117 & 0.242 $\pm$ 0.020 & Astrometry \\ 
        2M16411273+4028257 & 3.581 $\pm$ 0.015 &  164.197 $\pm$ 0.222 & 0.211 & 0.932 $\pm$ 0.109 & 0.567 $\pm$ 0.321 & Astrometry \\ 
        2M16462555+3934567 & 2.623 $\pm$ 0.010 & 2.766 $\pm$ 2.51E-04 & 51.807 & 1.015 $\pm$ 0.132 & 0.195 $\pm$ 0.016 & SB1 \\ 
        2M17284156+5614329* & 5.124 $\pm$ 0.012 &  676.150 $\pm$ 14.748 & 0.308 & 0.708 $\pm$ 0.083 & 0.242 $\pm$ 0.043 & Astrometry \\
        \hline
    \end{tabular}
    \caption{Orbital parameters for AGGC systems with Gaia solutions, and the two systems from \citet{corcoran2021} with orbits determined from their APOGEE RV curves. We provide the parallax, orbital period, maximum radial velocity shift (\drv), mass of the MS star ($M_{\rm MS}$  -- from \texttt{StarHorse}), minimum mass of the WD ($M_{\rm WD}$), and the method by which the orbital solution was obtained. For the systems with astrometric solutions, the value for the mass given in the `min $M_{\rm WD}$ column refers to the best-fit companion mass. Systems marked with * might not be WD+MS binaries, but contaminants from chromospheric activity in single MS stars.}
    \label{tab:gaia}
\end{table*}

In addition to \drv, some systems in AGGC also have orbital information that we use in our analysis.  
Eleven WD+MS systems in AGGC have an orbital solution from Gaia data release 3, either from astrometry or from radial velocities (see Table~\ref{tab:gaia}). For the single-lined spectroscopic binaries (SB1), measurements of the semi-amplitude of the primary RV ($K$) are also available. These measurements can be combined with primary mass ($M_{\rm MS}$) estimates from the Bayesian isochrone-fitting code \texttt{StarHorse} \citep{queiroz2023} (also in Table~\ref{tab:gaia}) to estimate the minimum mass of the WD. For systems with only an astrometric solution, we can estimate $K$ and consequently $M_{\rm WD}$ using the Gaia estimates for the Thiele-Innes parameters. The error on $M_{\rm WD}$ is estimated using Monte Carlo (MC) error propagation. We also include two systems for which orbital parameters could be derived from their APOGEE RV curves \citet{corcoran2021}.

\section{Binary population synthesis with COSMIC}\label{sect:cosmic}

Binary population synthesis (BPS) codes are widely used in the study of populations of single and binary stars. 
BPS calculations rely on tunable prescriptions to describe how the binary components and the orbit itself react to interactions via tides, wind mass loss, or Roche-lobe overflow which can remain stable on radiative or thermal timescales or lead to dynamically unstable common envelope evolution. `Rapid' BPS codes implement single star evolution based on a grid of single star models computed by \cite{Pols1998} through a series of analytic formulae that are fit to the models \citep[the `Hurley fits' ---][]{hurley2000}.

In this work, we use COSMIC\footnote{\href{cosmic-popsynth.github.io}{cosmic-popsynth.github.io}}, a Python-based, open-developed BPS code that has been modified to incorporate several updates to binary interactions and contains modules for the creation and analysis of large binary populations \citep{breivik2020}. While COSMIC has several parameters that control the evolution of binaries of all mass ranges, our main interest in this work are WD+MS binary systems.
We explore three key binary evolution parameters in our COSMIC simulations, and otherwise apply the default assumptions as described in COSMIC v3.4.10:
\begin{itemize}
    \item $\beta$, which sets the overall fraction of donor material that is accreted, with the rest being lost from the binary system ($\beta$ = 0.5 assumes 50\% accretion efficiency) \citep{belczynski2008}. It regulates the amount of mass accreted during stable Roche-lobe overflow. Material that is not accreted is isotropically re-emitted from the vicinity of the accretor. Values range from 0 to 1. 
    \item $\alpha$, the common envelope efficiency parameter, which scales the efficiency of transferring orbital energy to the envelope to eject it, as $E_{\rm bind, i} = \alpha (E_{\rm orb, f} - E_{\rm orb, i})$, \citealp[Eq. 71 of][]{hurley2002}. 
    $\alpha = 1$ indicates that all of the orbital energy of the system was used to unbind the envelope; larger values assume all of the orbital energy and energy from another, unspecified source, were used. 
    \item \qcf, which selects the prescription used to determine critical mass ratios (\qc, the ratio of donor mass/accretor mass) for the onset of unstable mass transfer and/or a common envelope during RLO depending on the evolutionary state of the donor. The prescriptions we use are described in Table \ref{tab:qcflag}.
\end{itemize}

\begin{figure}
    \centering
    \includegraphics[width=\hsize]{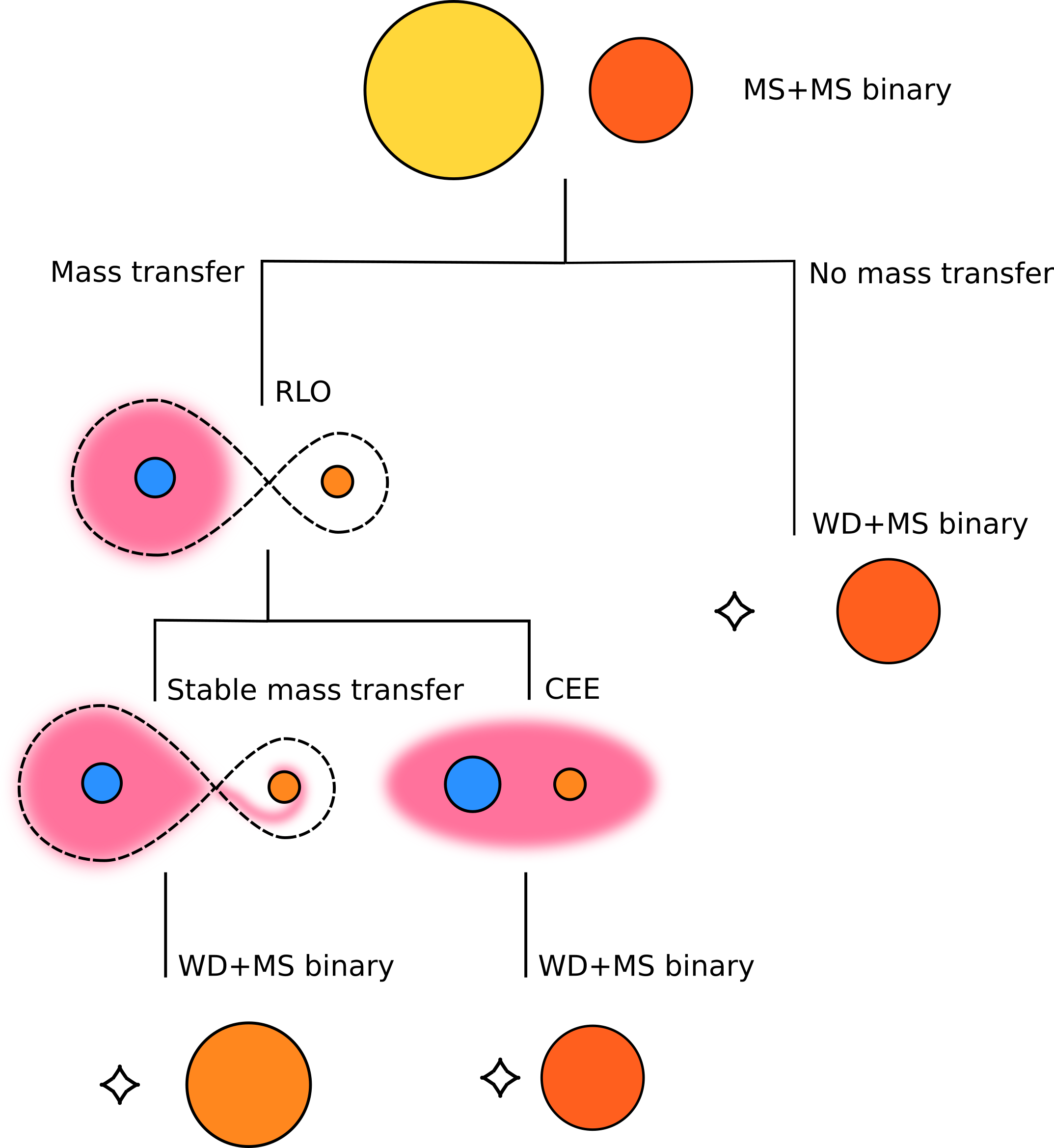}
    \caption{Schematic of the possible evolutionary paths through which a binary system can go to become a WD+MS binary. The system can either not interact, or go through RLO mass transfer, which can in turn be stable or unstable, leading to a common envelope phase. The imprints of their evolution can be seen in the final masses and orbital period of the systems. }
    \label{fig:diagram}
\end{figure}

WD+MS systems can form via multiple evolutionary channels. Figure~\ref{fig:diagram} illustrates these different evolutionary pathways. In our models, wide WD+MS systems can be formed without binary interactions, but the formation of close WD+MS binaries requires RLO interactions to shrink the orbit. Systems that become WD+MS binaries with periods lower than $\sim$1000 days were likely subject to a mass transfer episode in their lifetime. A given binary will traverse a single path based on the chosen binary evolution model and initial binary properties, but large simulated populations allow us to map the most common pathways that are taken for different parts of the initial binary population under a given set of assumptions that make up our binary evolution model. 
Changing the binary evolution assumptions within COSMIC changes the recipes for the interactions that these systems go through, significantly modifying both specific evolutionary paths and the characteristics of the final simulated WD+MS population. 

Whether a system goes through only stable mass transfer or through a phase of unstable mass transfer (such as CEE) depends not only on the initial separation of the system, but also on the amount of mass accreted during Roche-lobe overflow ($\beta$), on the response of the orbit to the mass transfer, and on the critical value for the mass ratio between the donor star and the accretor ($q_{\rm crit}$). This critical value in turn depends on the evolutionary state of the donor at RLO. We explore different values for \qc for key evolutionary states, namely the first ascent, asymptotic, and thermally pulsing giant branch stages. We define the \qcf parameter to indicate which prescription a given model uses for its \qc values. In COSMIC, there are different prescriptions for \qc across the evolution of the star, as detailed in Table \ref{tab:qcflag}. The \qcf = 1 prescription follows \citet{claeys2014}, where Eq. 1 is
\begin{equation}\label{eq:c}
    q_{\rm crit} = 2.13 \times \left[ 1.67 - x + 2\left( \frac{M_{c1}}{M_{1}}\right)^5\right]^{-1} ,
\end{equation}

\noindent with $x \approx 0.3$ at solar metallicity \citep{hurley2002}. The other prescriptions are the same, but vary the \qc of key evolutionary states: \qcf = 2 has \qc = 2.0 for {asymptotic giant branch (AGB)} stars, \qcf = 3 has \qc = 2.0 for {first ascent giant branch (FGB)} stars, and \qcf = 4 has \qc = 2.0 for {thermally pulsing giant branch (TAGB)} stars. 

If CEE happens, the fate of the system depends on the efficiency of the ejection of the envelope ($\alpha$). COSMIC uses the $\alpha \lambda$ prescription of \citet{hurley2002}, where $\lambda$ represents how bound the envelope is to its stellar core, i.e., a multiplicative factor in $E_{\rm bind}$. This parameter is calculated at each evolutionary state, following the description in Appendix A of \citet{claeys2014}. We do not explore different prescriptions for $\lambda$ directly since the fits provided in \citet{claeys2014} are drawn from the same stellar evolution models as the Hurley fits. However, $\alpha$ and $\lambda$ are degenerate, and thus our parameter $\alpha$ can be interpreted as a conjugate parameter $\alpha \lambda$.
If ejection is very effective ($\alpha \geq$ 1), the system has higher chances of surviving without merging, forming binaries with wider orbital periods than compared to less effective ejection ($\alpha \leq 1$). The effects of $\alpha$ increasing or decreasing, however could also be interpreted as increases or decreases in $\lambda$ such that the envelope of a CEE donor is either more or less bound to the core. 

\begin{table*}[]
    \centering
    \begin{tabular}{c c c c c}
        \hline 
        \hline
        \multirow{2}{*}{Evolutionary state of donor} & \multicolumn{4}{c}{\qc value for each \qcf} \\
         & 1 & 2 & 3 & 4 \\
        \hline
        MS, < 0.7 ${\mathrm{M}_\odot}$ & 0.695 & 0.695 & 0.695 & 0.695  \\
        MS, > 0.7 ${\mathrm{M}_\odot}$ & 1.6 & 1.6 & 1.6 & 1.6  \\
        Hertzsprung Gap & 4.0 & 4.0 & 4.0 & 4.0  \\
        First Ascent Giant Branch (FGB) & Eq. \ref{eq:c} &  Eq. \ref{eq:c} & 2.0 & Eq. \ref{eq:c}  \\
        Core Helium Burning & 3.0 & 3.0 & 3.0 & 3.0  \\
        Early Asymptotic Giant Branch (AGB) & Eq. \ref{eq:c} & 2.0 & Eq. \ref{eq:c} & Eq. \ref{eq:c}  \\
        Thermally Pulsing AGB (TAGB) & Eq. \ref{eq:c} & Eq. \ref{eq:c}  & Eq. \ref{eq:c} & 2.0  \\
        \hline
    \end{tabular}
    \caption{Values for $\rm q_{crit}$, the critical mass ratio for the onset of unstable mass transfer ($M_{\rm donor}$/$M_{\rm accretor}$), defined  according to the evolutionary state of the donor star for each \qcf. Eq. \ref{eq:c} from \citet{claeys2014}. }
    \label{tab:qcflag}
\end{table*}

\begin{table}[]
    \centering
    \begin{tabular}{c|c c c}
        \hline 
        \hline
        Model name & $\alpha$ & $\beta$ & $\rm q_{set}$\\
        \hline
        A[1, 2, 3, 4] & 0.3 & 0.0 & 1, 2, 3, 4  \\
        B[1, 2, 3, 4] & 0.3 & 1.0 & 1, 2, 3, 4 \\
       C[1, 2, 3, 4] & 1.0 & 1.0 & 1, 2, 3, 4 \\
        D[1, 2, 3, 4] & 5.0 & 0.0 & 1, 2, 3, 4 \\
        E[1, 2, 3, 4] & 5.0 & 1.0 & 1, 2, 3, 4 \\
        \hline
    \end{tabular}
    \caption{Overview of the parameters and names of 20 COSMIC models used in this work. Models are referred to by their names (A2, D3, etc.) throughout the text.}
    \label{tab:models}
\end{table}

We simulate several populations, each initialized with {$5 \times 10^6$} binary systems. 
These populations are initialized with primary masses drawn from the initial mass function from \cite{kroupa2001}, secondary masses drawn from a uniform distribution between $M_2=0.08\,M_{\odot}$ and the primary mass, a uniform eccentricity distribution, and the period distribution from \cite{raghavan2010}. The populations all have solar metallicity and a uniform star formation history over the last 10 Gyr. \citet{thiele2023} finds that using a metallicity-dependent initial binary fraction in COSMIC can have a considerable effect on the size of the final population of WD binaries (in particular double WD systems). While this effect can be significant, we do not explore metallicity in this work since the impact of different assumptions for mass transfer far outweigh the incorporation of metallicity-specific close binary fractions.  

We created 20 models varying $\alpha$, $\beta$, and \qcf parameters according to Table \ref{tab:models}. The models can be separated in three categories: low, middle and high $\alpha$. Models A and B have $\alpha = 0.3$ and $\beta=0.0$ and $1.0$, respectively; model C has $\alpha = 1.0$ and $\beta = 1.0$ (commonly assumed in population synthesis studies); and models D and E have $\alpha = 5.0$ and $\beta = 0.0$ and $1.0$, respectively. Within each family (A, B, C, D and E), we also explore the four prescriptions of \qcf, labeled 1 - 4 for reference (see Table~\ref{tab:qcflag}). 
Table \ref{tab:cosmicsystems} presents an overview of the number of systems in our simulations that end up as WD+MS binaries, double WDs (DWDs), WD+non-MS star, and mergers. In this work, we focus only on the WD+MS population.

{The result of each simulation, given our constant star formation history, is a `present-day' population of binaries, from which we gather the WD+MS pairs and apply observational selection cuts to produce a synthetic catalog of WD+MS systems. As the AGGC has a UV magnitude cut, we impose the same cut on our simulated populations 
We use the temperature of the two stars to estimate their combined black body emission in the far- and near-UV (FUV and NUV filters from GALEX) and remove systems with (FUV–NUV) < 5 from our synthetic catalogs \citep{anguiano2022}. We also remove systems that are WD+MS binaries, but are currently still interacting, as such systems are not expected to be present in the AGGC. Table \ref{tab:cosmicsystems} shows the number of systems present in each COSMIC model before and after these criteria are imposed on the population. 

The WD+MS sample of AGGC might be subject to contamination by non-binary MS stars with strong chromospheric activity in the UV, particularly in the range (FUV–NUV) < 2. However, as the effects of chromospheric activity in the AGGC are not fully characterized, we do not include this magnitude cut in our modeling. There is also bias in both APOGEE and Gaia towards brighter MS targets; however, correcting our models for this bias is non-trivial and beyond the scope of our analysis since it would require assigning positions in the Galaxy and associated extinction and crowding. This choice leads to the inclusion of potential both lower and higher mass MS companions, where lower mass companions may not be bright enough for APOGEE or Gaia to see, while higher mass companions may be rarer and thus not present in the data. We discuss the impact of this choice below in Section~\ref{sec:discussion}.

To compare the COSMIC models directly to the AGGC data, we calculated the \drv distributions for each simulated population. The periods, mass ratios and eccentricities of the binaries in each COSMIC model were used as the input to a Monte Carlo (MC) code \citep{badenes2018,mazzola2020} that generates random inclinations (in cosine space) and arguments of pericenter to project RV curves on the plane of the sky. From these RV curves, the MC code draws a random initial phase and samples RV values for a given number of epochs and distribution of cadences (i.e., the times at which these RV ``measurements'' were taken for each system). We used the cadence distribution from APOGEE data release 17 to accurately reflect the cadence in our simulated populations. Finally, the \drv was calculated for each sparsely sampled RV curve, resulting in a distribution of \drv for the synthetic catalog. As this is an MC code, the \drv distribution can be sampled many times, i.e., many sets of RVs can be obtained for the same system, given its inclination, argument of pericenter and time lag change at each realization. Thus, we can also estimate the uncertainty in the \drv distribution for our COSMIC models. For our comparison between data and models, we created 100 MC realizations of of the \drv distribution for each of our 20 COSMIC models.

\section{Results}\label{sect:results}

\subsection{Period distribution in the COSMIC models}

As an illustrative example of the way mass transfer affects orbital separations, Figure~\ref{fig:porbs} shows the initial and final period distribution in models C[1,2,3,4] ($\alpha = 1.0$ and fully conservative mass transfer, with different prescriptions for \qc, following Table~\ref{tab:qcflag}), for the systems that are present-day WD+MS binaries. 
The top row shows the initial orbital period distribution for the WD+MS progenitors colored by their evolutionary pathway. For illustrative purposes, we also show initial orbital period distribution for the entire population in dark gray, arbitrarily normalized. The period distribution of the WD+MS population is shown in the bottom row with the same color scheme. The WD+MS population period distribution varies widely from model to model. This indicates that different mass transfer stability assumptions are more important than the initial separation when determining the post-interaction separation of WD+MS binaries. 


\begin{figure*}
    \centering
    \includegraphics[width=0.95\linewidth]{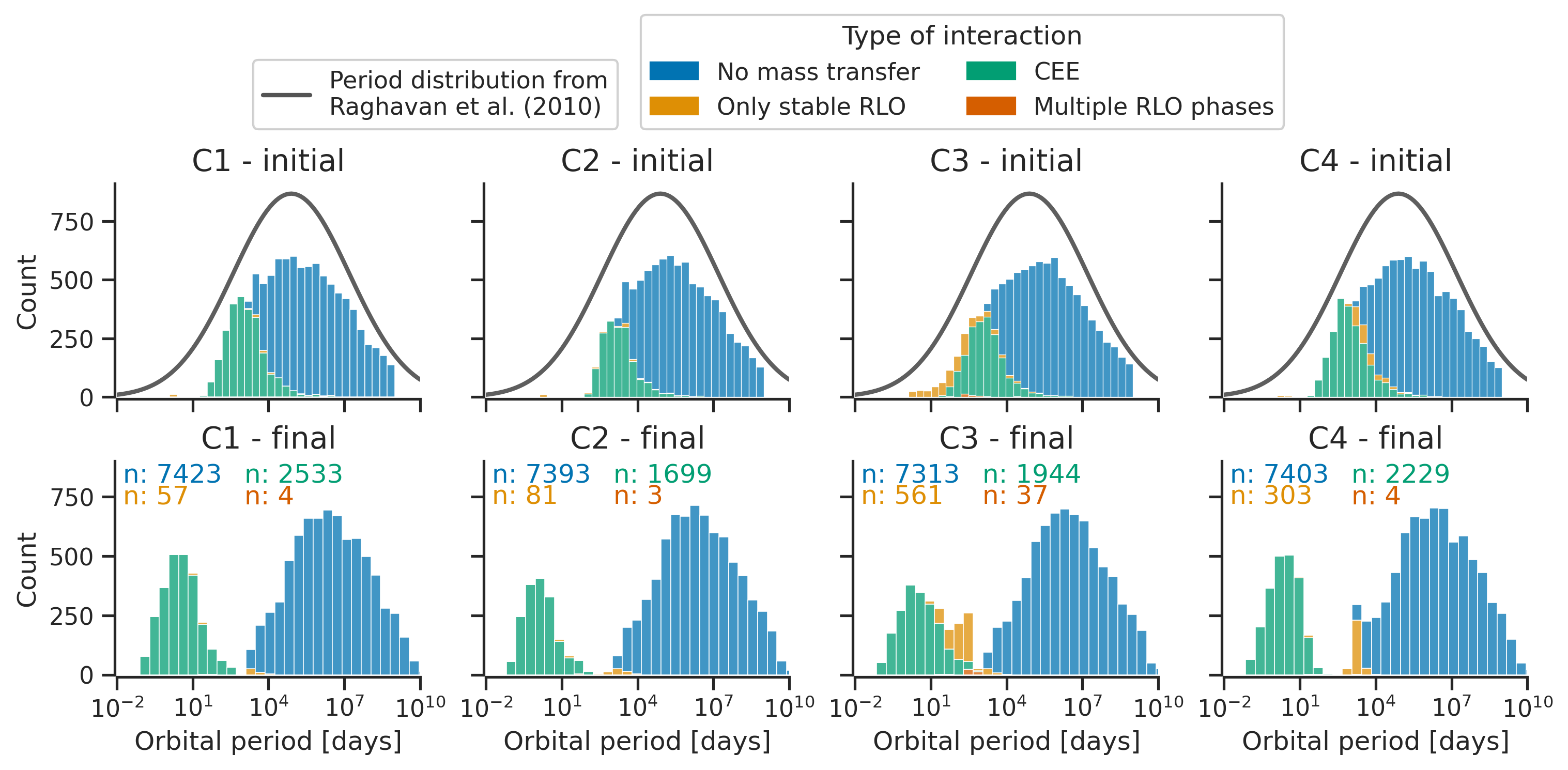}
    \caption{Initial (top rows) and final (bottom rows) period distribution for the WD+MS binaries in models C1, C2, C3 and C4 ($\alpha = 1.0$ and fully conservative mass transfer, with different prescriptions for \qc, following Table~\ref{tab:qcflag} -- colored histograms). The colors indicate the type of interaction the system undergoes to end up as a WD+MS binary. The numbers in the bottom rows indicate the number of systems in the final distribution according to type of interaction.
    The dark gray line in the top plots shows the normalized period distribution for the entire simulated population (following \citealt{raghavan2010}), including systems that do not end up as WD+MS binaries, but as double WDs, MS+MS binaries, mergers, and WD binaries with post-MS stars. All models are initialized with the same period distribution. The final period distribution of the WD+MS binaries are quite different between the models, indicating a stronger dependence on mass transfer stability assumptions rather than initial orbital separation.}
    \label{fig:porbs}
\end{figure*}

Figure~\ref{fig:porbs} also shows the relative number of systems in each COSMIC model broken down by evolutionary channel: no mass transfer, stable RLO mass transfer, and CEE, with each channel colored by the type of interaction the system went through during its evolution. The majority of systems, for all models, become a WD+MS system without ever interacting (blue histogram); these systems have the longest orbital periods, as it is necessary that they be in wide orbits to avoid interaction in the first place.
In all models, the systems that end up with the shortest periods are those that undergo CEE. 
Our choice of \qcf affects both the number of WD+MS binaries that undergo stable RLO or CEE as well as their period distributions. Generally, stable mass transfer leads to longer periods. However, binaries which undergo CEE but have longer initial periods can have present day periods that are similar to binaries which go through stable mass transfer but originate with shorter periods.

Our models with \qcf = 1 follow the widely used prescription of critical mass ratio for unstable mass transfer (\qc) from \citet{claeys2014}. In models with \qcf = 2, we explore a prescription where mass transfer occurring while the donor is in the AGB phase is more prone to stability, as the \qc value for this evolutionary stage is higher. For models with \qcf = 3, mass transfer is more stable during the FGB phase, and for models with \qcf = 4, during the TAGB phase (see Table~\ref{tab:qcflag}). As Figure~\ref{fig:porbs} indicates, there is little variation in the WD+MS binary period distribution between models with \qcf = 1 and 2 within a given model family, but discrepancies are more significant for \qcf = 3 and 4. In particular, \qcf = 3 produces a much higher fraction of stable systems than the other prescriptions, and a lower mean orbital period. A value of \qcf = 4 also yields a larger number of stable systems in all model families, but the period distribution is wider, with a high density of systems around 1000 days, and some outliers with shorter periods, of the order of tens of days. This is because the WD+MS binary progenitors for \qcf = 3 which undergo stable mass transfer originate in shorter orbital periods compared to the stable mass transfer WD+MS binary progenitors in \qcf = 4.

The accretion limit, $\beta$, has a limited impact on the final period distributions. There is a slight decrease in the number of systems that remain stable between the A and D models ($\beta$ = 0.0 --- i.e., non-conservative accretion) and the B, C and E models ($\beta$ = 1.0 --- i.e., fully conservative). 
The final period distribution of the interacting systems becomes less extended in all models with $\beta = 1.0$ compared to models with $\beta = 0.0$. 
Models C have a larger value for the common envelope ejection parameter ($\alpha = 1.0$) than models A and B, meaning that all the gravitational binding energy of the system is used to expel the envelope of the donor. Models D and E have $\alpha = 5.0$, where an additional source of energy is used to remove the envelope. Because ejection is more easily achieved, more systems manage to avoid merging, increasing the overall size of the population of WD+MS binaries formed via CEE. The number of systems that merge in our COSMIC simulations increases the more conservative the mass transfer is, as the orbits shrink more effectively due to the conservation of angular momentum.

Figure~\ref{fig:histporbs} shows histograms of the period distribution for only the systems that undergo mass transfer (whether stable or unstable) in each model where the color indicates different model choices. The top panel shows the dependence of the period distribution on \qcf, keeping $\alpha$ and $\beta$ fixed. There is a noticeable region with a low density of systems in the period distribution between $\sim$200 and $\sim$1000 days in all model families, with the clear exception of model A3. This absence of a gap is true for all models in family 3 (A3, B3, C3, D3 and E3) and models with high $\alpha$ (families D and E -- see middle and bottom panels of Figure~\ref{fig:porbs}). The period gap is produced by the boundary between stable and unstable mass transfer. The systems that cross this boundary have the necessary conditions (mass, radius and separation) to enter a CEE phase, where mass and angular momentum transfer tighten the orbit of the system on a dynamical timescale. The differences in the size and exact position of the gap in period space depend, therefore, on the mass ratio threshold for unstable mass transfer, \qc. In models with \qcf = 1, 2 and 4, FGB stars have a \qc set by Eq. \ref{eq:c}, which gives values in the range of 0.7 - 1.5. In contrast, models with \qcf = 3 have this value set to 2.0, allowing more systems to end mass transfer without going into CEE, pushing the mean period of these systems to lower values, and closing the gap in orbital period. 

The vertical lines in Figure~\ref{fig:histporbs} represent the median of the period distributions for the products of stable mass transfer (dashed lines) and CEE (dash-dotted lines), with values shown in Table \ref{tab:cosmicsystems}. The median period does not change significantly for the products of CEE between models A and B. In model family C, the median of the final periods of systems that undergo CEE increases when compared to low-$\alpha$ models, with systems moving to wider periods in the 2-3 day range and populating the period gap, with the exception of model C2. This shows that the increased stability in the mass transfer during AGB phase (\qcf = 2) does not significantly impact the evolution of WD+MS binaries that form via CEE. High-$\alpha$ models D and E show an even clearer trend of CEE products with larger periods, with the mean for all \qcf jumping to tens of days, varying little among themselves, and closing the period gap present in lower $\alpha$ models. The middle panel of Figure~\ref{fig:histporbs} shows the dependence on $\alpha$ and clearly depicts this trend of increasing mean period of post-CEE systems. In the bottom panel of Figure~\ref{fig:histporbs} only $\beta$ varies. Its effect is very limited, and the changes in the median of the period distributions are small.

For RLO products, the effect of both $\alpha$ and $\beta$ is not nearly as relevant as the effect of \qcf, with only slight variations in the median of the period distribution. Models A[1,2] and B[1,2] and models D[1,2] and E[1,2] show a trend of increasing median values indicating that their periods increase with $\beta$. Models [A,B]4 do not show the same behavior, with the median period instead \textit{decreasing} slightly with $\beta$. The increase in period mean is particularly noticeable between models D1 and E1, going from 92 to 1754 days, due to the double peaked nature of the distribution of RLO products in model D1 (see this same model in the second to last panel of Figure~\ref{fig:porbs}). The overall number of RLO products increases greatly in the 3 family of models. The period gap is filled up by the WD+MS binaries as stable mass transfer products for these models, which show only slight variations between each other, and thus do not seem to depend strongly on the value of either $\alpha$ or $\beta$.

\begin{figure}
    \centering
    \includegraphics[width=\hsize]{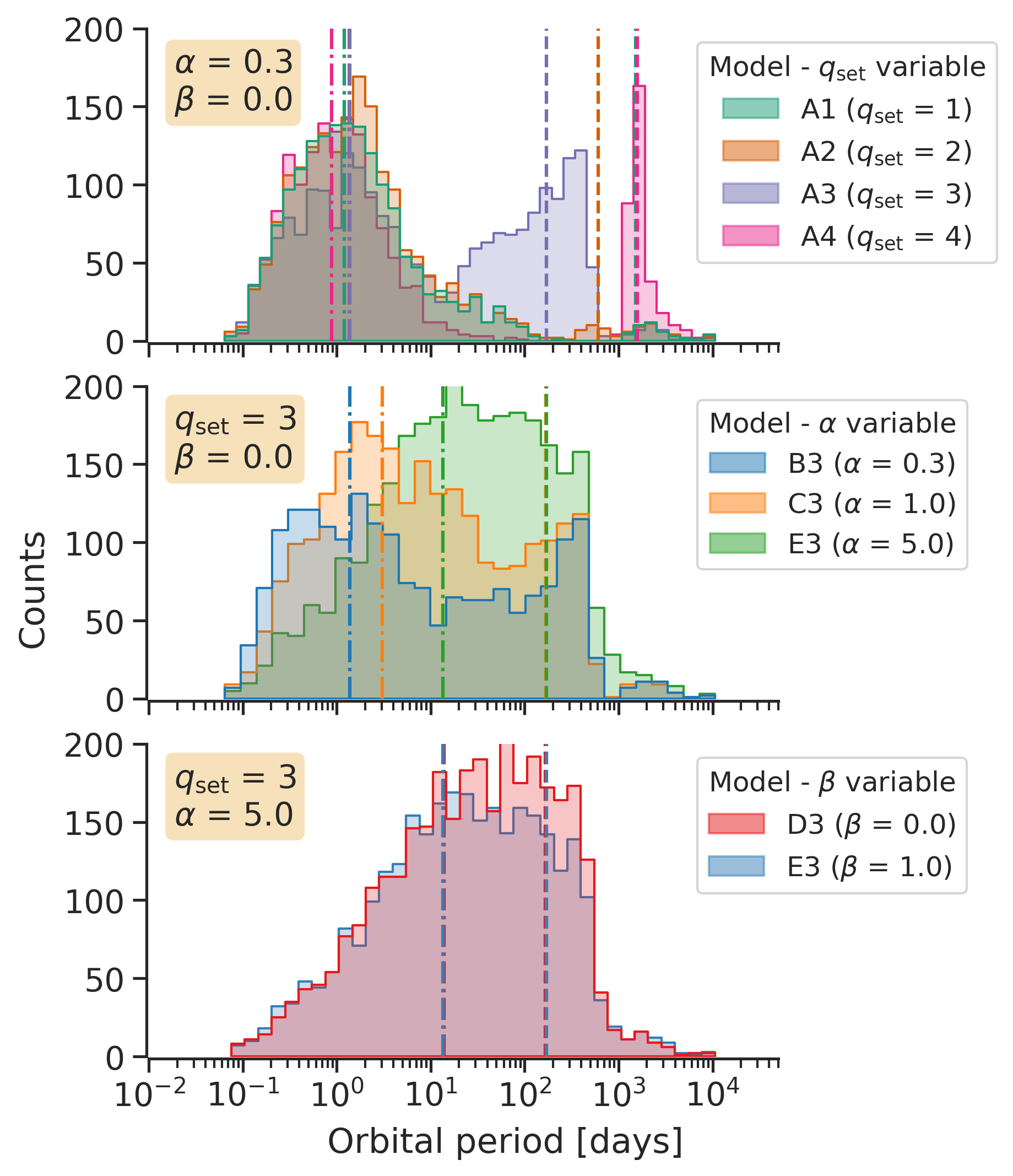}
    \caption{The orbital period distribution of the WD+MS binaries that went through mass transfer, either stable or unstable, in our COSMIC models. The top panel shows the variation in the period distribution due to \qc; middle panel shows the same, but due to $\alpha$, and the bottom panel, due to $\beta$. The dashed lines show the median value of the period distribution considering only the stable mass transfer systems. The dashed-dotted lines show the median period for the CEE systems. Values for the medians are shown in Table \ref{tab:cosmicsystems}. The largest changes in the period distribution occur when \qc = 3, where the first ascension giant branch stars tend to have stable mass transfer. The main effect of increasing $\alpha$ is creating a post-CEE systems with larger periods, while the effect of $\beta$ is nearly negligible.}
    \label{fig:histporbs}
\end{figure}

\subsection{Models vs. Observations: \drv distribution, periods and WD masses}\label{sect:drvm}

Our COSMIC models yield the final distributions of periods, eccentricities, and masses of the simulated systems.
As described in Section \ref{sect:cosmic}, we estimate the \drv distribution for each COSMIC model by supplying this information, plus a series of time lags, to an MC code. Figure~\ref{fig:rainbow} shows how the period distribution generated by a COSMIC model (in this case, model A3) translates into a \drv distribution using this method. The left panel shows the period distribution of model A3, colored by logarithmic period ranges. The stacked histogram on the right plot shows the \drv distribution resulting from 100 realizations of the MC code.
The largest values of \drv correspond to the shortest period binaries, but the correlation between orbital period and \drv\ becomes less straightforward for lower \drv\ values, as \drv depends not only on period, but on the inclination of the system and on the timing of the observations: a system seen face-on will show no RV variation, and a system with a 50 day period will have a small \drv if its RV measurements are all taken within a span of only a couple of days.

\begin{figure}
    \centering
    \includegraphics[width=1\linewidth]{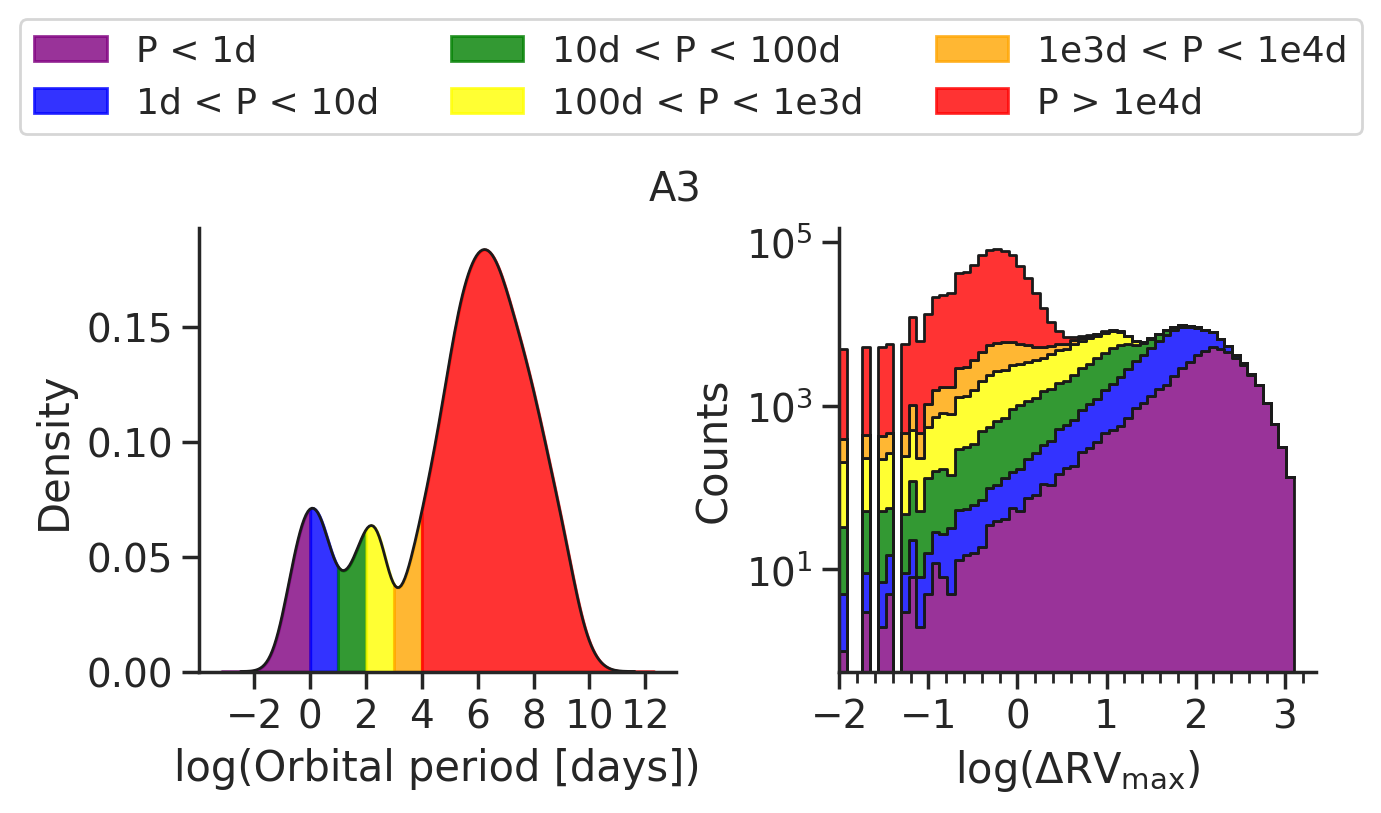}
    \caption{The distribution of orbital periods (left panel) and \drv (right panel) for model A3. There is an anti-correlation between period and \drv, but it is possible for short period systems to have small \drv, as \drv has a strong dependence on the  inclination, eccentricity, and the time lag between radial velocity observations. Larger \drv are more unambiguously correlated with short period systems.}
    \label{fig:rainbow}
\end{figure}

\begin{figure*}
    \centering
    \includegraphics[width=0.9\linewidth]{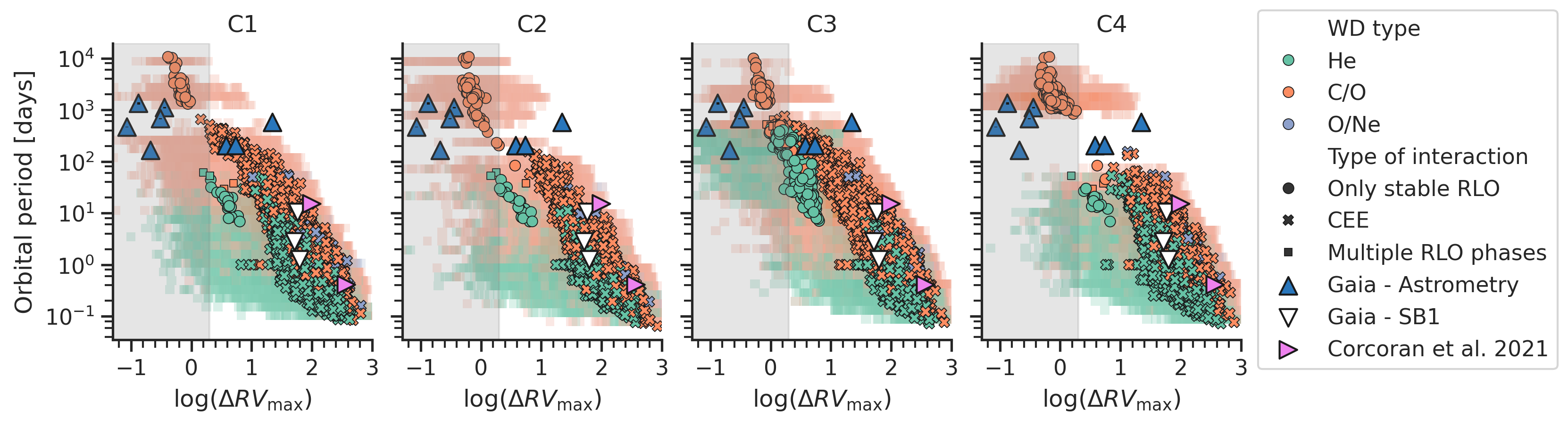}
    \caption{The orbital periods and \drv for the C family of models ($\alpha$ = 1.0, $\beta$ = 1.0) and observational data (triangles). Colors indicate the type of WD formed after mass transfer, and the markers indicate the type of mass transfer the system went through. Non-interacting systems are not shown. The gray shaded region indicates the region below \drv = 2km/s, where RV uncertainties dominate. The background squares show 2D $\log_{10}$-scaled histograms of all 100 MC realizations of the \drv distribution for each binary system in the model. The points for \drv are the mean of the \drv distribution for each binary in each simulated population derived from our MC sample. Model C3 is the best match to the data, showing a more continuous period/\drv distribution. }
    \label{fig:drvmax_porb_data}
\end{figure*}

\subsubsection{Systems with orbital solutions from Gaia}\label{sec:gaiasystems}

Figure~\ref{fig:drvmax_porb_data} shows the correspondence between orbital period and \drv for models in family C ($\alpha$ = 1.0, $\beta$ = 1.0), overlaid with the values for the AGGC systems that have orbital solutions from Gaia (large blue triangles for astrometric orbits, and large white inverted triangles for single line spectroscopic binaries -- from Table~\ref{tab:gaia}). 
The underlying 2D histograms show the $\log_{10}$-scaled 
spread of the \drv sampling over 100 MC realizations. The small symbols represent the mean \drv values for each of the binary systems, with the shape of the symbol indicating the type of mass transfer interaction (circles for RLO, crosses for CEE, and squares for multiple RLO phases).
The region where \drv < 2 km/s is shaded gray in the plot. In this regime, individual \drv measurements
can be affected by RV uncertainties \citep{badenes2018} that are not considered in our modeling \citep[see ][Figure 3, Table 1 for a discussion]{mazzola2020}. A version of this plot with all models is shown in Fig~\ref{fig:drvmax_porb_data_all} in Appendix~\ref{sec:AppA}.

For values of \drv above a few km/s the AGGC systems with Gaia orbital solutions show a continuous relation of periods and \drv. 
In contrast, as discussed in Section \ref{sect:cosmic} above, several COSMIC models show a gap (a region of where no, or only relatively few WD+MS systems are formed) in the orbital period distribution. As seen in Figure~\ref{fig:drvmax_porb_data}, model C1 has a small gap around 1000 days, model C2 has a decrease around 300 days, and model C4 has a large gap between 100 and 1000 days. As already evidenced by the period distributions in Figure~\ref{fig:histporbs}, the effect of \qcf is extremely relevant in determining the size, position, and intensity of this gap. 

It is clear that model C3 covers the same parameter space as the systems with Gaia orbits, while models C1, C2 and C4 struggle to reproduce the data. We note that we have few systems with Gaia orbits, and that the 5 systems with \drv < 2 km/s had their \drv obtained from APOGEE observations taken within a few-day window, leading to a lower estimated \drv than expected from their periods.

Figure~\ref{fig:drvmax_porb_data} also indicates the composition of the WD in our models. {In COSMIC, whether the final WD is a Helium (He), Carbon/Oxygen (C/O), or Oxygen/Neon (O/Ne) WD is determined by its final mass as enforced by the amount of nuclear burning allowed in the stellar core, taking into account any mass transfer episodes.} He WDs have masses lower than 0.45 $M_\odot$, C/O WDs between 0.45 and 1.05 $M_\odot$, and O/Ne above 1.05 $M_\odot$. A noted difference between models with \qcf = 3 and the rest is that in these models, the gap is filled with WDs of different origin and composition. For instance, in model C1 the only way the COSMIC simulations create systems with orbits of 100 to 1000 days is via CEE, and the resulting WD is always a C/O WD; on the other hand, systems can form in that period space in model C3 via either RLO or CEE, showing a mix of He and C/O WDs. Thus, establishing the mass and composition of the WD in the observed systems in this region of the period/\drv plot could help determine which model provides the best description of the true WD+MS population.

The AGGC systems with orbital solutions from Gaia have estimates for the mass of MS star. Using these estimates and the RV semi-amplitude of the primary ($K$) measured by Gaia, we can infer the minimum mass of the WD companion. {With this value, the composition of the WD can be estimated. While the correlation between WD mass and composition is not straightforward for products of binary interaction, we tentatively confront this estimated WD composition of the observed sample to the models. }

Figure~\ref{fig:pdmass_all} compares the estimated mass of the MS star and WD type for the systems shown in Table~\ref{tab:gaia} to models [A,C,D,E]3, favored by the \drv comparison (see Sect. \ref{sect:drvm}). 
As shown in Table~\ref{tab:gaia}, the estimate of the mass (and thus type) of the WD depends on the inclination of the orbit. As such, the WD types shown by the colors in Figure~\ref{fig:pdmass_all} relate to the minimum WD mass. Thus, systems that have a minimum WD mass corresponding to a He WD might not in fact have a He WD when the inclination is accounted for; these data are colored yellow in the Figure. Considering these effects, models A3 and D3 provide the best qualitative agreement with the data. 

There is a clear trend in the location of the He WD companions in the period range of 10-1000 days to shift to higher MS companion masses for increased $\beta$ (A3 to B3, D3 to E3). The He WDs in this period range are all stable mass transfer products (see Figure~\ref{fig:drvmax_porb_data}). As the mass accretion becomes more efficient, less massive (He) WDs can only be formed if the MS donor does not lose too much of its mass in the mass transfer. 
Only models A3 and D3 (both with non-conservative mass accretion) can reproduce the He WDs in that period region found in the data. We note that the nature of these WDs as He WDs is an estimate based on the available data. Confirming the mass of these WDs will provide constraints to the mass transfer efficiency for these models. 

Additionally, all models predict a high number of low mass, short period ($P < 10$ days) binaries that are not present in the data. The cause for this discrepancy is likely the detection limit of APOGEE, since these low mass MS stars are faint, our sample size is small, and this bias is not accounted for in our modeling. 
Increasing the number of WD+MS systems with well measured masses, orbital periods, and temperatures will facilitate a more robust statistical comparison between models and data in the future. For example, the WD binaries pathway survey, beginning in 2016 \citep{parsons2016}, has confirmed both individual WD+MS binaries and populations obtained through both spectroscopic and astrometric survey data with spectroscopic follow up studies.

\begin{figure*}
    \centering
    \includegraphics[width=\linewidth]{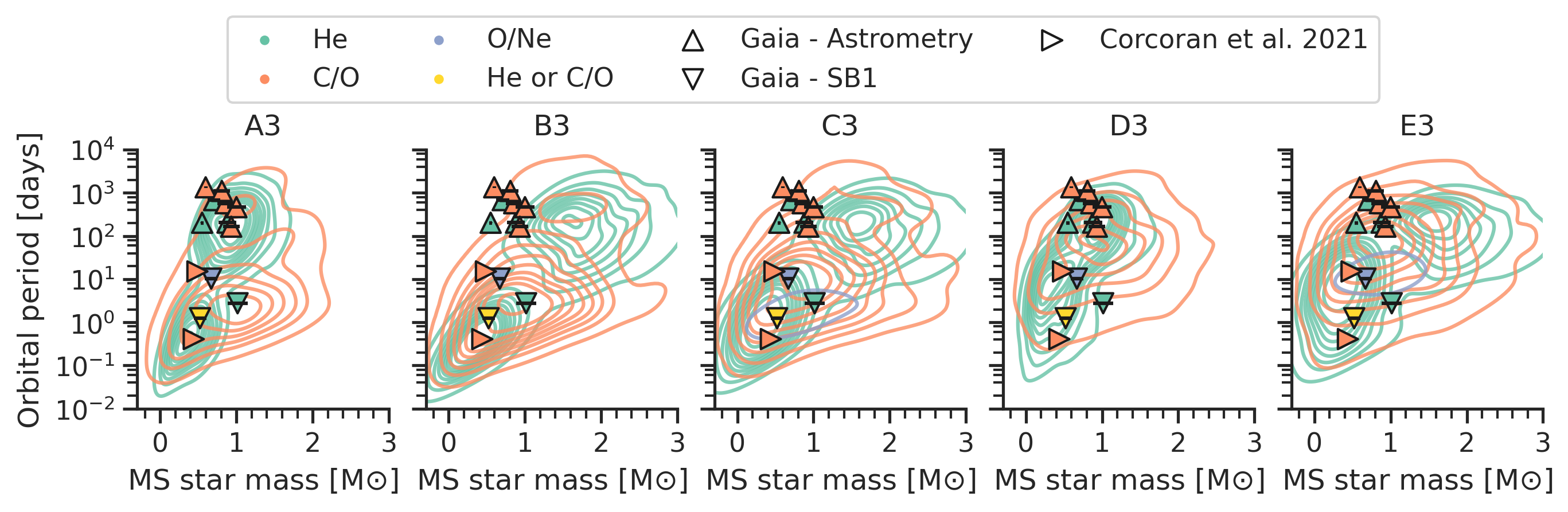}
    \caption{Orbital period versus mass of the MS component of the WD+MS binary systems. The background KDE contours represent the model (from left to right, A3, B3, C3, D3 and E3). The triangles represent observational data from Table~\ref{tab:gaia}. The colors indicate the type of WD companion: teal for He, orange for C/O, purple to O/Ne, and yellow for WDs that can be either He or C/O depending on the inclination of the system. Models A3 and D3, both with $\beta$ = 0.0, are the only models that can reproduce the large period ($P$ > 100 days), low mass MS+He WD systems. }
    \label{fig:pdmass_all}
\end{figure*}

\subsubsection{Systems with only \drv\ from APOGEE}

We also compare the \drv distributions predicted by our COSMIC models with the entire AGGC sample of WD+MS systems using histograms and cumulative distribution functions (CDF). Histograms can be complicated to interpret due to binning and cutoff effects (see Appendix~\ref{ap:cutoffs}), but two features are almost always present in our simulations regardless of these details: a peak near log(\drv/(km/s))$=1.8$ and a dip near log(\drv/(km/s))$=1$}. 
Figure~\ref{fig:histdrvm} shows the \drv histograms for the 100 MC realizations for models [A,B,C,D,E]3 (light green blocks) and the AGGC \drv data (dots) where the width of the blocks represents the 90\% confidence region for each bin. The \drv distribution histograms cover the range from 5 to 1000 km/s. The low end cut on 5 km/s is motivated by uncertainties in the \drv measurements from APOGEE, and the fact that the large period systems present in our models that dominate in low \drv are likely not present in the AGGC data, as wider binaries are more difficult to detect.

With this cut, the number of AGGC data points is 98. Thus, we divided the distribution in $\sqrt{98} \sim$ 10 log-spaced bins for our analysis. 
The data peaks close to log(\drv) = 2.0, suffers a decrease, and plateaus for log(\drv) < 1.5.

\begin{figure}
    \centering
    \includegraphics[width=0.7\linewidth]{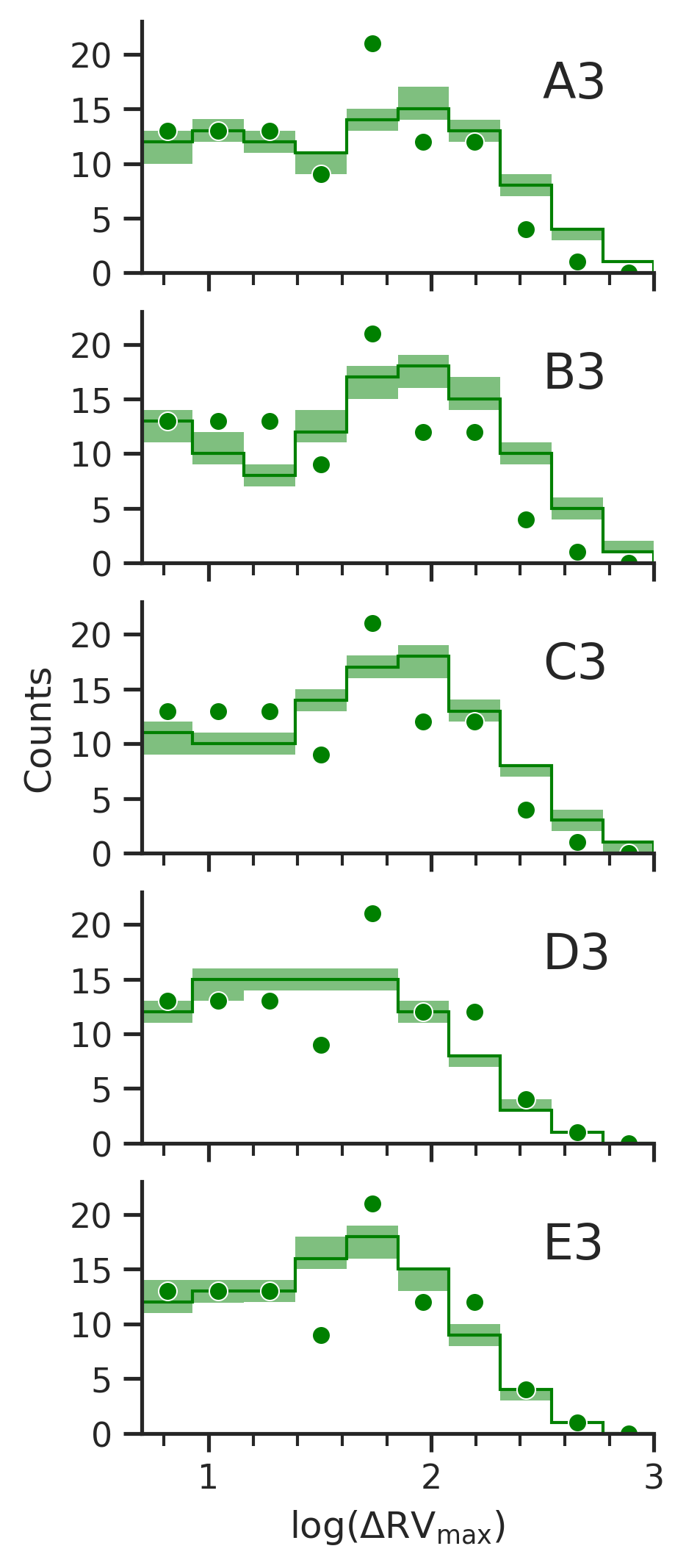}
    \caption{\drv histograms for 100 MC realizations for COSMIC models with \qcf = 3 (light green blocks) and the AGGC \drv data (dots); the width of the blocks represents the 90\% confidence region for each bin. The \drv distribution histograms cover the range from 5 to 1000 km/s, with 10 bins. }
    \label{fig:histdrvm}
\end{figure}

\begin{figure*}
    \centering
    \includegraphics[width=0.9\linewidth]{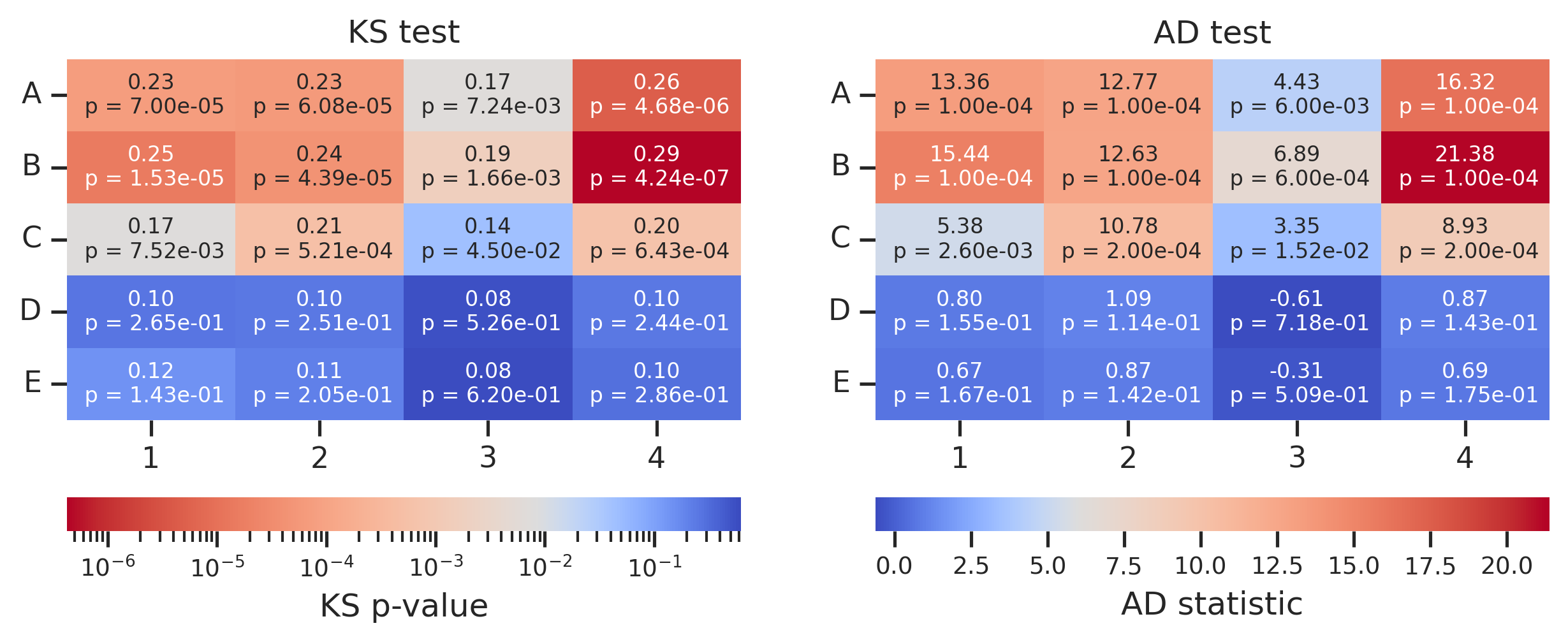}
    \caption{KS and AD statistics for the comparison of the \drv distribution of COSMIC models to data. Histograms are shown in Figure~\ref{fig:histdrvm_full}. Model to data comparisons that yeld KS p-value smaller than 0.05 and AD statistic lower than 6.546 cannot be discarded as being sampled from the same underlying distribution. Models with $\alpha = 5.0$ (families D and E) and models [A,D]3 are the best matches to the data.}
    \label{fig:ksad}
\end{figure*}

We also used the two-sample Kolmogorov-Smirnov (KS) and the Anderson-Darling (AD) statistical tests on the \drv cumulative distributions of the models and data. Figure~\ref{fig:histdrvm} shows the results of these statistical tests comparing the data with the mean CDF of all MC realizations for each model. Figure~\ref{fig:ksad} shows the results of both tests for all COSMIC models.

The null hypothesis that the data and models were sampled from the same underlying distribution is rejected by the AD test at the a certain level when the AD statistic is larger than a given critical value. The values are and levels are 0.325 (25\%), 1.226 (10\%), 1.961 (5\%), 2.718 (2.5\%), 3.752 (1\%), 4.592 (0.5\%), 6.546 (0.1\%). Therefore, models with an AD statistic larger than 6.546 strongly reject the hypothesis that the two samples come from the same underlying distribution, while models with a statistic lower than 0.325 cannot reject the null-hypothesis. For models [A,B][1,2,4], C[2,4] and model B3, the hypothesis can be rejected. Models [D,E][1,2,4] can be rejected only at the 10\% level. Model C3 is rejected at the 2.5\% level, model A3 at the 1\% level, and models [D,E]3 cannot be rejected. On the KS test, all models in families D and E and model C3 have p-values larger than 0.05, so the null hypothesis that the data and the models could be drawn from the same distribution cannot be rejected. 

In addition to the KS and AD test, we also considered different measurements of statistical distance between the data and the models. We also vary the number of bins and the lower end cutoff of the \drv distribution to guarantee that our results are independent of these factors. The results of these tests are shown in Appendix \ref{ap:cutoffs}, Figure~\ref{fig:stat_distances}.

As the results of all statistical tests performed indicate, we conclude that the only models that cannot be outright rejected statistically are models with high $\alpha$ (families D and E), and models A3 and C3, where the critical mass ratio for the onset of unstable mass transfer, \qc, for FGB stars is fixed at 2.0. This value is higher than the one assumed in the widely used \qc prescriptions of \citet{hurley2002} and \citet{claeys2014}. Our higher value indicates that stars that fill the Roche-lobe in this evolutionary phase are more prone to undergo stable mass transfer. We note, however, that a larger sample would be required to draw more robust statistical comparison to our BPS models.

\section{Discussion}\label{sec:discussion}


Determining the stability of mass transfer in binaries is crucial to understand the formation channels and subsequent evolution of post-RLO systems. Given the complexity of this problem, the classical approach has been to estimate the response of mass transfer based on simplifications to the structure of the donor star and the assumption that its response to mass transfer is adiabatic. Thus, donors with radiative envelopes would tend towards stability, while giants with convective envelopes would quickly turn unstable \citep{hjellming1987}. Many works in the past couple of decades have demonstrated that these simplifications underestimate the stability of mass transfer for many mass ranges. From the observational side, \citet{leiner2021}, for instance, finds that the number of blue stragglers found in clusters on Gaia DR2 cannot be reproduced by COSMIC using the \qc prescriptions of \citet{hurley2002} or \citet{hjellming1987}.

Theoretical studies have also found indications that the instability criteria need to be reevaluated. \citet{pavlovskii2015}'s models of mass transfer find higher stability for giant donors when the response of the superadiabatic layers of the star to mass loss is considered in stellar evolution models. The mass loss sequences of \citet{ge2010, ge2020}, based on more realistic models of stellar interiors, yield mass–radius relations that map the threshold of thermal to dynamical time scale mass loss for a wide range of mass ratios. Their results indicate that donors on the RGB and AGB tend to be more stable (i.e., harder to begin rapid mass transfer and go into CEE) than previous models indicated. 

The work of \citet{temmink2023} delves deep into the issue of mass transfer with post-MS donors. Using the stellar evolution code MESA \citep[][and others]{paxton2011, jermyn2023}, they follow the evolution of the donor until it fills its Roche lobe without imposing a fully adiabatic response on the star, which prevents a too rapid expansion of the convective envelope. Their results indicate that giant donors undergo stable mass transfer for a wider range of mass ratios than the widely used prescription of \citet{hurley2002}. However, rapid BPS codes continue to use these prescriptions to determine the threshold of stability of their systems due to both software development lag times and lack of information on stellar interior structures due to the use of fitting formulae.

Comparing the \drv and period distributions of our COSMIC models with the AGGC data shows that several models are unable to reproduce the general trends seen in the data. In Figure~\ref{fig:drvmax_porb_data}, models [A,B,C][1,2,4] predict a gap in the period distribution that is not seen in the data. The results of the \drv distribution (Figure~\ref{fig:histdrvm}) also rule out these models as good representations of the population of WD+MS binaries of AGGC. On the other hand, models with \qcf = 3 and models with $\alpha = 5.0$ appear to more accurately capture observed features in the data. While our findings rely on simple assumptions for the stability of mass transfer, our flexible approach arrives to the same conclusions found in detailed stellar evolution modeling. Similar tests applied to larger datasets in the future may be able to connect theoretical predictions to observations in order to apply constraints on mass transfer across a wide range of the binary population parameter space.

The distribution of WD and donor masses with period offers even more insight, although the masses are more uncertain than the \drv and period estimates. Of the models favored by the \drv distribution, only A3 and D3 cover the same parameter space as the data in donor masses and period for He WDs with periods larger than 100 days. This result indicates a preference for low $\beta$, or less conservative mass transfer (Figure~\ref{fig:pdmass_all}). By increasing $\alpha$, binary systems with He WDs are formed in a wider range of periods, reaching as high 100 days, also creating an overall larger population of WD+MS binaries. Even so, the data show He WDs with periods on the order of several hundreds to nearly 1000 days, which can only be reproduced in our models by increasing \qc for the onset of instability in first ascent giant stars (\qcf = 3).


High $\alpha$, or extremely efficient envelope ejection in CEE, is preferred by our results, as it provides a \drv distribution that matches the high \drv end of the AGGC distribution (see Figures~\ref{fig:histdrvm} and \ref{fig:histdrvm_full}). In practice, this means that the AGGC data require that a larger fraction of post-CEE systems (which dominate the high end of the \drv distribution --- see Figure~\ref{fig:drvmtype}) have larger periods. Since models with lower $\alpha$ over-predict the number of short periods, a large $\alpha$ is preferred.
Still, $\alpha$ = 5.0 is significantly higher than the `standard' value of 0.3 \citep[i.e.,][]{zorotovic2010, scherbak2023}. An $\alpha$ > 1 requires that all of the binding energy plus an additional source of energy be used to expel the envelope. This extra energy would likely come from thermal energy, recombination energy from the expansion of the envelope \citep{zorotovic2014, nandez2016, invanova2018}, and nuclear energy due to interaction between the two stars \citep{podsiadlowski2010}. 

Whether or not any additional energy is necessary to create post-CEE WD+MS binaries with periods on the order of tens of days is currently a matter of debate. The catalog of WD+MS candidate binaries compiled by \citet{shahaf2023} from the third Gaia data release offers insight. Follow up work in \citet{yamaguchi2024A} and \citet{yamaguchi2024B} seek to validate the orbits of WD+MS binaries in these catalogs for massive O/N WDs (the former) and for more standard C/O WDs (the latter). In both studies, the formation pathways for the observed systems are considered by simulating the evolution of single stars up to and through the thermal pulses on the AGB. Both papers find that the observed wide orbital periods, between 100 to 1000 days, can be produced by WD progenitors which initiate CEE with low envelope binding energies due to the highly evolved donor star.

In the case of the O/N WD progenitors, the inclusion of recombination energy leads to the observed periods \cite{yamaguchi2024A}. In the case of the C/O WD progenitors, long period post-CEE WD+MS binaries can be formed without the need for the additional recombination energy, even for low values of $\alpha$. However, while recombination energy is not strictly required to explain the long period post-CEE WD+MS binaries, their models show that it dominates the internal energy in the outer envelope of the donor during the TAGB phase, and therefore should play a significant role in the envelope ejection. Including recombination energy also produces long period binaries for a wider range of initial parameters.

Another important point is that the envelope-structure parameter, $\lambda$, changes significantly when recombination energy is included in the modeling. This is clearly illustrated in see their Figure 13e of \citet{yamaguchi2024B} 
during the TAGB phase for models with and without recombination energy. During the thermal pulses, $\lambda$ oscillates around values near 0.5 for the model with no recombination, while the value goes from -1 (i.e., the envelope is unbound) up to 5 when recombination is considered.


In COSMIC, we use the $\alpha \lambda$ prescription for the CEE phase. As in BSE, COSMIC sets $\lambda$ based on the calculations of \citet{claeys2014}. While we are only exploring different values for $\alpha$ between our models, these two parameters are degenerate: the same effect could be achieved by modifying $\lambda$ instead. Thus, it is more accurate to interpret our results for $\alpha$ as the combined effect of $\alpha \lambda$ as well as potential additional energy sources. Our results indicate that either CEE donors have much lower envelope binding energies or that recombination energy is necessary to explain the distribution of \drv. In particular the higher end, dominated by short period binaries, which are over predicted by our low $\alpha$ models. This result is consistent with \citet{yamaguchi2024B}'s interpretation that recombination energy does play a significant role in the TAGB phase, making them more efficient at forming WD+MS binaries with wider periods. The significant effect of the recombination energy on $\lambda$ during the TAGB phase shown by \citet{yamaguchi2024B}'s models indicates that its implementation in BPS codes are unlikely to capture the full range of $\lambda$ values that occur during the late-stage evolution of WD progenitors. Given the degeneracy with $\alpha$, it is possible that discrepancies between values of $\alpha$ found in different BPS studies are actually due to changes in $\lambda$ between (and during) different evolutionary phases of the donor.

The recent work of \citet{torres2025} uses  
inverse population synthesis techniques to reconstruct the evolution of WD+MS systems. Their modeling uses the $\alpha \lambda$ prescription, but they only use orbital to unbind the envelope which implicitly enforces a null internal energy contribution. This means that $\alpha$ is limited to 1 in their results. Applying their algorithm to a sample of 30 eclipsing WD+MS binaries, they find a strong correlation between the mass of the WD progenitor and $\alpha$, and that a single value of $\alpha$ cannot reconstruct all systems in their sample. We propose that this correlation in $\alpha$ is caused by the degeneracy between $\alpha$, $\lambda$, and the mass of the WD progenitor, as these parameters are connected via the binding energy of the envelope and orbital separation at the onset of the first RLO. The limitation of $\alpha\leq 1$ and assumed prescription for $\lambda$ requires a change in the mass of the WD progenitor to account for the required changes in envelope binding energy. Our simulations choose the same prescription for $\lambda$, but do allow for $\alpha>1$ (and find a slight preference for $\alpha=5$). While we don't account for possible correlations between mass and $\alpha$, the assumptions for $\lambda$ do account changes in binding energy due to radial expansion of the WD progenitor. If, as suggested by \citet{yamaguchi2024A,yamaguchi2024B}, our applied assumptions for $\lambda$ over predict the envelope binding energy, the correlations found by \citet{torres2025} may be accounted for naturally with upgraded assumptions to stellar physics.

In another study, \citet{belloni2024A} used the population synthesis code BSE \citep{hurley2002} to simulate post-CEE binaries containing O/N WDs to compare with massive WD binaries reported in \citet{yamaguchi2024A}. \citet{belloni2024A} found that wide O/N WD binaries could be formed assuming only efficient envelope ejection, with no extra source of energy required. This is achieved by allowing the WD progenitors to enter CEE as a highly evolved thermally pulsing AGB (TAGB), where the donor star radius is larger and there has been significant mass loss due to winds than at the AGB phase. In this case, the amount of orbital energy required to eject the envelope is less due to lower binding energies.  
The most similar model to their work is B1, which uses $\alpha = 0.3$, conservative mass transfer, and a critical mass ratio based on \citet{claeys2014}, the same as BSE's default value for giant stars. However, since our simulated WD+MS populations are dominated by He and C/O WD binaries, it is difficult to make a direct comparison to \citet{belloni2024A}. Future studies which consider larger simulations may be able to provide a more comprehensive comparison, but are currently outside the scope of this paper.

Finally, we consider two important caveats to this work. First, our results assume that AGGC contains a pure sample of WD+MS binaries. However, as mentioned in Section~\ref{sect:data}, the catalog might be contaminated by single MS stars with high chromospheric activity, which are incorrectly assigned as WD+MS binaries by their excess in FUV. If indeed this is the case for a significant part of our dataset, it could significantly bias our results. Future studies larger complementary datasets could either uncover this bias or help validate our findings.

We also note that we only considered one assumption for how mass that is not accreted in stable mass transfer is treated. Future studies could explore different angular momentum loss assumptions where higher rates of angular momentum loss would lead to shorter periods after RLO, thus potentially increasing the effects of stable mass transfer in our 3 family of models. However, we expect this effect to be less dominant than our assumptions for the stability of mass transfer and CEE since the majority of the WD+MS population is formed through CEE.

\section{Conclusion}\label{sec:conclusion}

In this work, we compare WD+MS binaries from the APOGEE-Galex-Gaia catalog with synthetic binary populations calculated with the BPS code COSMIC. Because the AGGC offers the distinct advantage of a large and homogeneous sample of WD+MS binaries with multiple high-quality RV measurements for each system, we used their \drv measurements (and the full orbital solutions for some systems) to calibrate the binary mass transfer parameters used in COSMIC. We explored a range of values for the common envelope ejection efficiency ($\alpha$), the accretion limit ($\beta$) and the critical mass ratios which define the onset of unstable mass transfer (\qc) used in COSMIC, which are also standard parameters in other BPS codes.    

Overall, model D3, best describes both the \drv distribution (Figures~\ref{fig:drvmax_porb_data} and \ref{fig:histdrvm}) and the measured WD masses in the AGGC data (Figure~\ref{fig:pdmass_all}). This suggests a preference for \qcf = 3 (defined in Table~\ref{tab:qcflag}) such that stable mass transfer during RLO for first ascent GB donor stars occurs mass ratios up to \qc = $2.0$, $\alpha$ = 5.0, and $\beta = 0.0$. We summarize these preferences below

\begin{itemize}
    \item \textit{More stable mass transfer on the FGB:} We can rule out models where FGB stars go into unstable mass transfer with \qc < 2 when $\alpha < 5$, as such models underpredict the number of small \drv/long period binaries (see models [A,B,C][1,2,4] in Figure~\ref{fig:histdrvm_full}).
    This result indicates that mass transfer from donor stars on the FGB is more stable than the widely used \qc prescription of \citet{claeys2014}. It also corroborates recent studies that find increased mass transfer stability in giant stars in both data and models \citep[i.e.,][]{temmink2023}.
    \item \textit{High $\alpha$:} In the \drv comparison, there is a preference for models with high common envelope ejection efficiency ($\alpha$), but the statistical significance is low, and models with \qcf = 3 and lower $\alpha$ should not be completely excluded. The preference for high $\alpha$ models indicates that the inclusion of some other energetic source during envelope ejection is necessary. However, the $\alpha$ parameter is degenerate with the envelope structure parameter $\lambda$. It might be the case that our results suggest that donor envelopes are less bound than the assumptions in COSMIC. Following \citet{yamaguchi2024B}, this may be the case during the thermally pulsing AGB phase, where $\lambda$ varies widely, and in a manner not accounted for in BPS codes.
    \item \textit{Low mass transfer efficiency}: When information on the minimum WD mass drawn from full orbital solutions is considered, models with low mass transfer efficiency ($\beta$) are preferred, as they are the only models that produce He WDs in the same parameter region as the data.       
\end{itemize}

We require more precise orbital solution and mass estimates to draw more robust conclusions on the mass transfer parameters of the population of WD+MS binaries. Comparing \drv distributions provides a general idea of the trends of the distribution in period space, but the addition of precisely-measured orbital periods of these observed systems to compare with the COSMIC models will allow direct constraints on mass transfer assumptions. More accurate WD mass estimates will also further constrain these assumptions. At present, the \drv distribution is the most effective indicator of the effect of CEE assumptions for our sample, but it is dominated by changes to our assumptions for mass transfer stability. Additional precise measurements for the periods and masses of WD binaries will allow more precise constraints that can help discriminate between values of $\beta$ (Figure~\ref{fig:pdmass_all}).

Our results showcase that large, homogeneous datasets can be used to constrain parameters rapid BPS codes which can be directly connected theoretical models for binary evolution. They also show that certain commonly used prescriptions, such as the $\alpha\lambda$ for CEE and the \qc of \citet{hurley2002}, should be revised to take into account the most recent results of stellar evolution modeling. In particular, careful treatment of envelope binding energies rather than variable envelope ejection efficiencies may be a direct route to understanding post-CEE WD binaries. 

The release of Gaia DR4 will provide orbital solutions to many more WD+MS binaries, and for other post-mass transfer systems as well. When these data are available, our comparison to BPS models can be reassessed for more accurate limits on the mass transfer parameters. Farther in the future, the Laser Interferometer Space Antenna (LISA) mission will lead to the detection and characterization of thousands of Galactic gravitational wave sources including close double WD binaries and other post interaction compact binaries. Comparing these data with BPS models will produce constraints on the mass transfer parameters for a wider range of initial conditions \citep[e.g.][]{thiele2023, delfavero2025}.

\begin{acknowledgements}
      This work was initiated at the Kavli Summer Program in Astrophysics 2023, hosted at the Max Planck Institute for Astrophysics. We thank the Kavli Foundation and the MPA for their support. ACR acknowledges support from ESO and its studentship program. ACR is grateful to Selma de Mink and Stephen Justham for helpful comments and discussion during the development of this project, and Alex Carciofi for continued support. ACR, KB and CB are also grateful to the AGGC collaboration for useful discussions on the manuscript. CB, BA, and SM acknowledge support from NSF-AST grant 2307865. KE acknowledges support from NSF grant AST-2307232.

\end{acknowledgements}

\appendix

\section{Summary tables and figures of full model set of COSMIC simulations}
\label{sec:AppA}

\begin{table*}[]
    \centering
    \begin{tabular}{c|cccccc|cc|cc}
        & Full pop. & & & & & & After cuts & & Median [days] \\
        Model & Total & WD+MS & WD+Other & DWD & Merger & Other & Total & WD+MS & RLO & CEE \\ 
        \hline \hline
         A1 & 65235 & 40070 & 1246 & 23389 & 3 & 527 & 14981 & 9128 & 1509.53 & 1.20 \\ 
        A2 & 66296 & 40294 & 1245 & 24186 & 7 & 564 & 15228 & 9202 & 603.44 & 1.36 \\ 
        A3 & 68554 & 42082 & 1411 & 24429 & 5 & 627 & 16175 & 9674 & 169.53 & 1.37 \\ 
        A4 & 67354 & 39927 & 1262 & 25416 & 211 & 538 & 15781 & 9129 & 1579.83 & 0.88 \\ 
        \hline
        B1 & 64405 & 39842 & 1208 & 22515 & 305 & 535 & 14951 & 9116 & 1835.94 & 1.27 \\ 
        B2 & 64988 & 39938 & 1233 & 22951 & 306 & 560 & 15174 & 9085 & 782.15 & 1.26 \\ 
        B3 & 66680 & 40255 & 1283 & 23528 & 1073 & 541 & 16607 & 9490 & 169.31 & 1.38 \\ 
        B4 & 66780 & 39725 & 1222 & 25118 & 180 & 535 & 15608 & 9044 & 1529.58 & 0.83 \\ 
        \hline
        C1 & 70776 & 43475 & 1320 & 23727 & 1561 & 693 & 17207 & 10017 & 1822.25 & 3.08 \\ 
        C2 & 64965 & 40056 & 1188 & 22861 & 304 & 556 & 15120 & 9176 & 1409.58 & 1.28 \\ 
        C3 & 73655 & 42984 & 1429 & 26126 & 2383 & 733 & 18599 & 9855 & 167.83 & 3.07 \\ 
        C4 & 71390 & 42921 & 1306 & 25518 & 1018 & 627 & 17322 & 9939 & 1524.73 & 2.37 \\ 
        \hline
        D1 & 78461 & 46406 & 1397 & 29999 & 26 & 633 & 18090 & 10551 & 92.14 & 12.26 \\ 
        D2 & 78605 & 46141 & 1412 & 30381 & 23 & 648 & 18249 & 10557 & 599.29 & 13.84 \\ 
        D3 & 82049 & 46898 & 1578 & 32864 & 26 & 683 & 19178 & 10626 & 165.28 & 13.86 \\ 
        D4 & 78416 & 45926 & 1428 & 30227 & 235 & 600 & 18209 & 10535 & 1590.89 & 10.78 \\ 
        \hline
        E1 & 77753 & 46152 & 1436 & 28625 & 832 & 708 & 18148 & 10499 & 1754.25 & 12.28 \\ 
        E2 & 77817 & 45838 & 1397 & 29455 & 433 & 694 & 18218 & 10468 & 907.67 & 13.23 \\ 
        E3 & 80798 & 45213 & 1480 & 32090 & 1311 & 704 & 19663 & 10407 & 169.33 & 13.51 \\ 
        E4 & 77889 & 45672 & 1416 & 29173 & 934 & 694 & 18444 & 10409 & 1513.65 & 10.01 \\ 
        \hline
    \end{tabular}
    \caption{Additional information on COSMIC models, showing the number of systems that end up as WD+MS binaries, WD+non-MS star, double WD (DWD), mergers, and other configurations (such as MS+MS binaries). We also show the number of WD+MS binaries that form our sample once the cuts in UV magnitude are applied. The last two columns show the values of the median of the period distribution of the final sample, separated by the time of mass transfer the system undergoes. These values are shown as the dotted and dashed lines in Figure~\ref{fig:histporbs}.}
    \label{tab:cosmicsystems}
\end{table*}

\begin{figure*}
    \centering
    \includegraphics[width=0.95\linewidth]{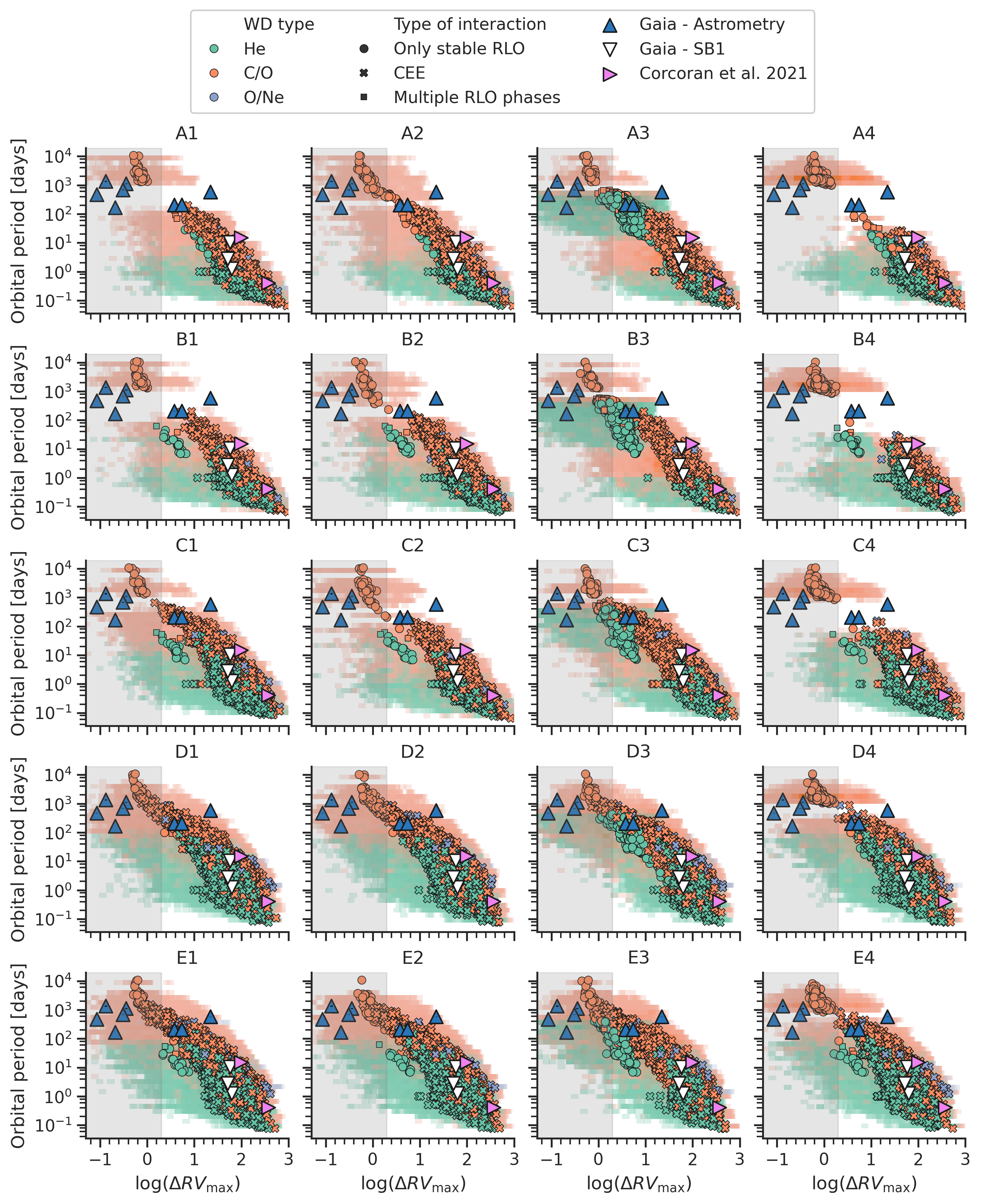}
    \caption{Orbital periods and \drv for all models and observational data (triangles). The colors indicate the type of WD formed after mass transfer, and the markers indicate the type of mass transfer the system went through. Non-interacting systems are not shown. The gray shaded region indicates our cut off in \drv. The background squares show 2D log-normed histograms of all 100 MC realizations of the \drv distribution for each binary system in the model. The points are the mean \drv distribution for each model.}
    \label{fig:drvmax_porb_data_all}
\end{figure*}

Figure~\ref{fig:drvmax_porb_data_all} is the expanded version of Figure~\ref{fig:drvmax_porb_data}, showing the period/\drv distribution for all models along with the AGGC systems with well determined Gaia orbital solution. Models A1, A4, B1, B4 and C4 all show defined gaps in their period/\drv relation, while models C2 and B2 show a less prominent gap feature. Generally, the location of the gap shifts with the values for common envelope ejection efficiency ($\alpha$) and fraction of mass accreted during stable mass transfer ($\beta$). Models in the C, D and E families ($\alpha$ = 1.0 and 5.0 --- three bottom rows) generally show smaller or no gaps when compared to models in the A and B families ($\alpha$ = 0.3 --- top rows), as the post-CEE systems tend to have larger periods owing to a more effective ejection of the common envelope. 

The accretion limit, $\beta$, also has an effect in populating the period range of 100--1000 days. Models A and B with a fixed $\alpha = 0.3$, but increasing values of $\beta$ show slight variations in the period distribution. 
In these variations, the B models produce He WDs with smaller \drv than the A models in the period range of 10-100 days. The B models also form fewer C/O WDs in this same period range when compared to the A models. A similar trend is also seen in high $\alpha$ models D and E, which have $\beta = 0.0$ and 1.0, respectively. The effect of modifying \qcf, however, is much stronger than that of the accretion limit, as already evidenced by the period distributions shown in Figure~\ref{fig:drvmax_porb_data}.  

For the 4 family models, where TAGB mass transfer is more stable, the gap spans a wider period range, from about 20 to 1200 days in models A4 and B4 to 100 to 1000 in model C4, closing completely in models D4 and E4. Finally, in models with \qcf = 3, or more stable FGB mass transfer, the gap no longer exists, being completely filled with systems resulting from mostly stable (circles and squares), but also unstable (crosses) mass transfer.

\begin{figure*}
    \centering
    \includegraphics[width=0.95\linewidth]{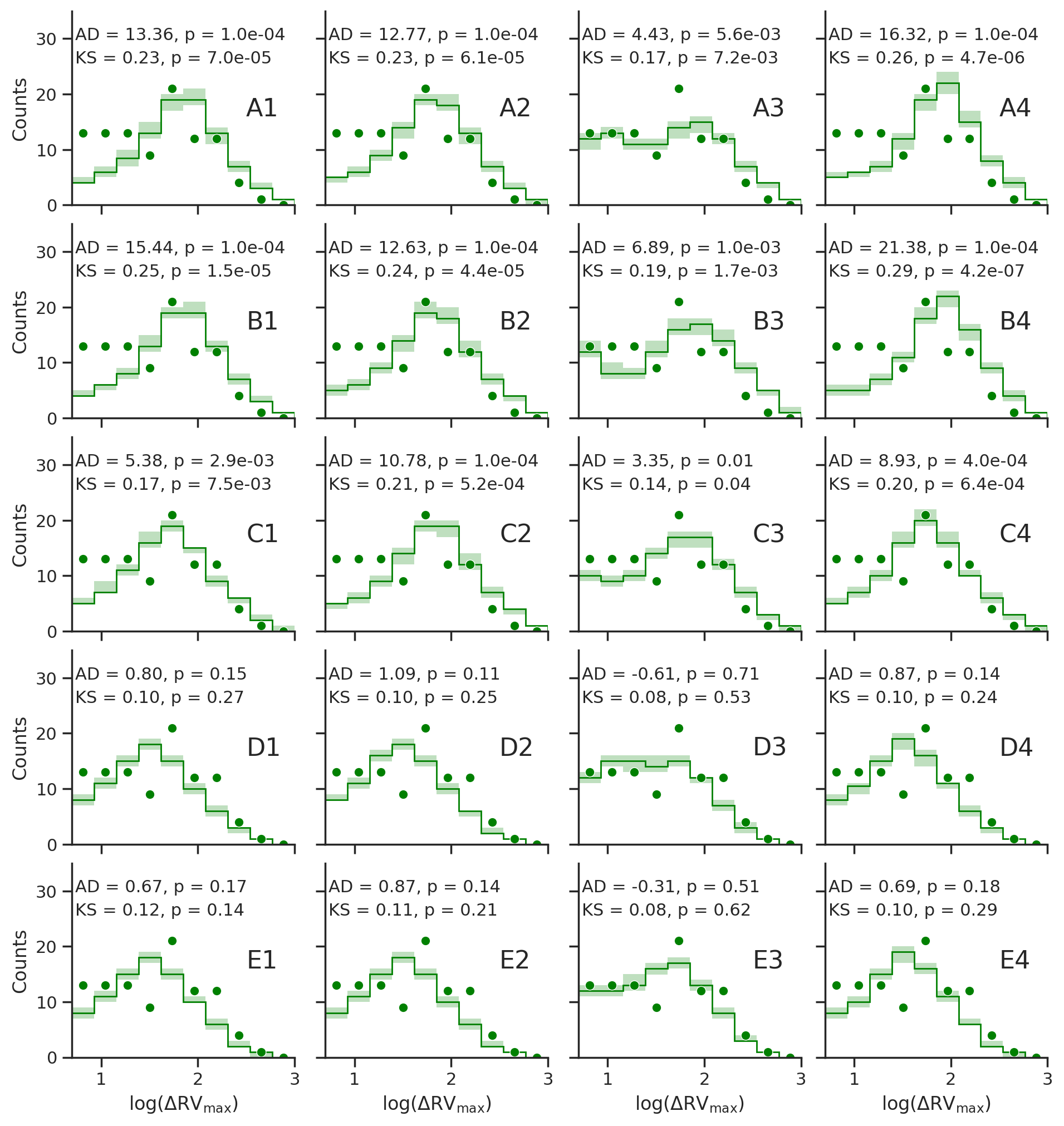}
    \caption{\drv histograms for 100 MC realizations for all 20 COSMIC models (light green blocks) and the AGGC \drv data (dots); the width of the blocks represents the 90\% confidence region for each bin. The \drv distribution histograms cover the range from 5 to 1000 km/s, with 10 bins. Models with low $\alpha$ and \qcf $\neq$ 3 ([A,B,C][1,2,4]) tend to underestimate the number os systems with lower \drv, and overestimate the systems with high \drv.}
    \label{fig:histdrvm_full}
\end{figure*}

Figure~\ref{fig:histdrvm_full} shows the histograms for all COSMIC models compared to the \drv measurements from the AGGC. Models in families 1, 2, and 4 cannot reproduce the low \drv plateau. The low and middle $\alpha$ models, [A,B,C][1,2], are able to reproduce the rapid rise in the in log(\drv) between 2 and 3 km/s, which high $\alpha$ models [D,E][1,2] underestimate. Models with \qcf = 4 behave similarly, but only model C4 captures the behavior of the high \drv peak; models [A,B]4 overestimate the number of systems, while models [D,E]4 underestimate them. Undoubtedly, models with \qcf = 3 are the ones that best describe the data, corroborating the results shown in Section \ref{sec:gaiasystems}, where these models are shown to reproduce the periods, \drv, and WD masses of the sample of WD+MS with Gaia orbital solutions.

\begin{figure*}
    \centering
    \includegraphics[width=0.95\linewidth]{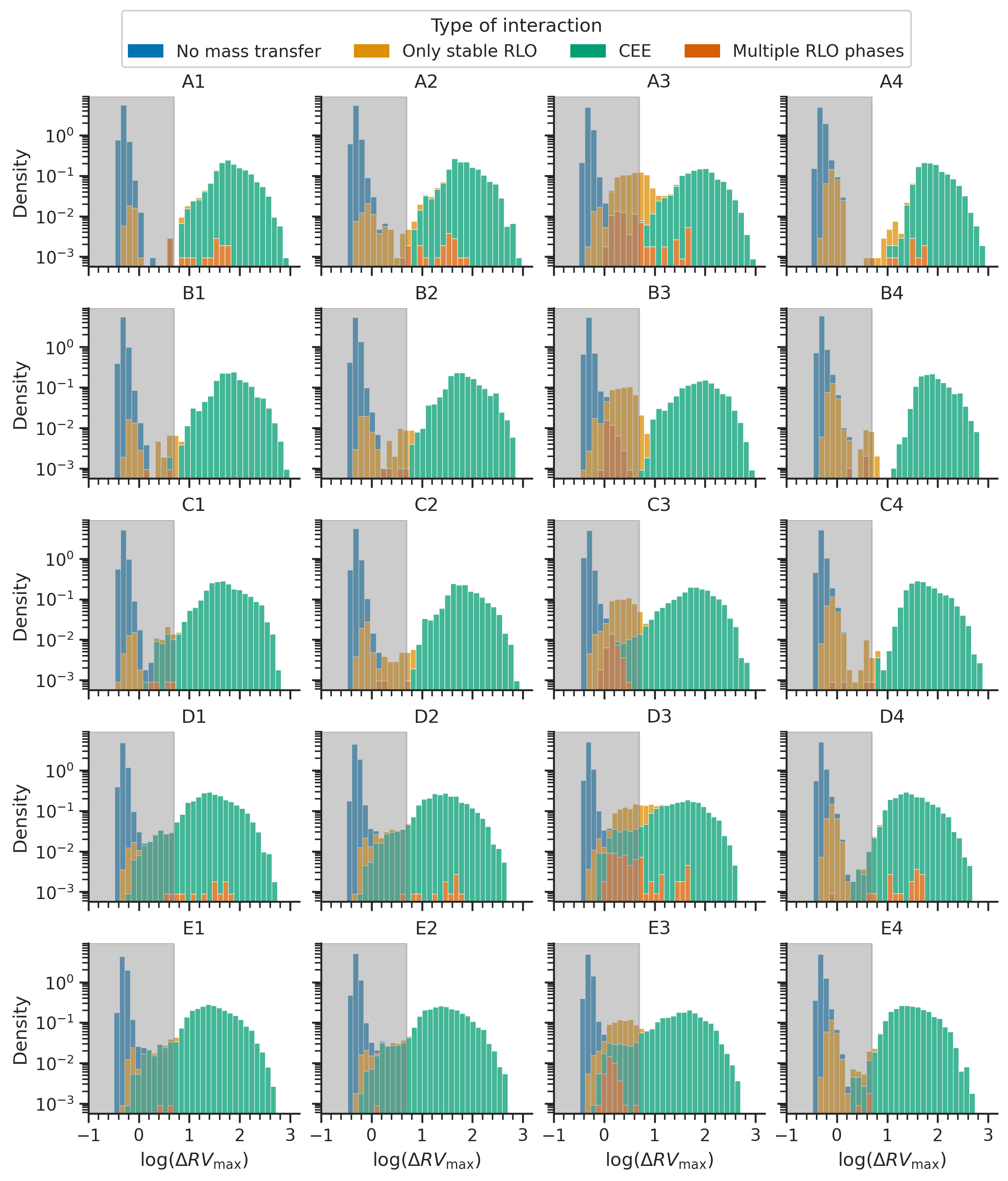}
    \caption{\drv distribution for all 20 COSMIC models, colored by the type of interaction the system suffered. The shaded gray region is the 5km/s cut. The \drv > 5km/s region is dominated by post-CEE WD+MS binaries.}
    \label{fig:drvmtype}
\end{figure*}

Figure~\ref{fig:drvmtype} shows the mean \drv distribution of all models (same as the full line in the histograms of Figs. \ref{fig:drvmax_porb_data} and \ref{fig:drvmax_porb_data_all}), colored by the type of interaction. The gray shaded region is our chosen \drv cutoff of 5km/s. The post-CEE systems dominate the distribution, as expected. This explains the strong dependence of our results on CEE parameters $\alpha$ and \qcf, and the much weaker effect exerted by $\beta$. The larger RV errors in the low \drv regime make the region dominated by stable mass transfer WD+MS binaries difficult to probe with only \drv. 

\section{Statistical tests, cutoff and binning effects}\label{ap:cutoffs}

We performed several statistical tests to quantitatively compare the \drv distributions of the AGGC data and our COSMIC populations. When comparing the histograms, we calculated a Poisson likelihood (as the sum of the log of the Poisson probability mass function in each bin); the total variation distance, defined as
\begin{equation}
    \Delta(D, M) = \frac{1}{2} \sum^N_{i=0}{|D_{i} - M_{i}|} ,
\end{equation}

\noindent where $D$ and $M$ are the counts of systems in each bin $i$; the Hellinger distance,
\begin{equation}
    H(D,M) = \frac{1}{\sqrt2 } \sqrt{\sum^N_{i=0}{\left( \sqrt{D_i} - \sqrt{M_i}\right)^2}} .
\end{equation}

We further consider the Manhattan, or city block distance,
\begin{equation}
    C(D, M) = \sum^N_{i=0}{|D_{i} - M_{i}|} ,
\end{equation}

\noindent and the Euclidean distance
\begin{equation}
    E(D, M) = \sqrt{\sum^N_{i=0}{|D_{i} - M_{i}|^2}}.
\end{equation}

\begin{figure*}
    \centering
    \includegraphics[width=\linewidth]{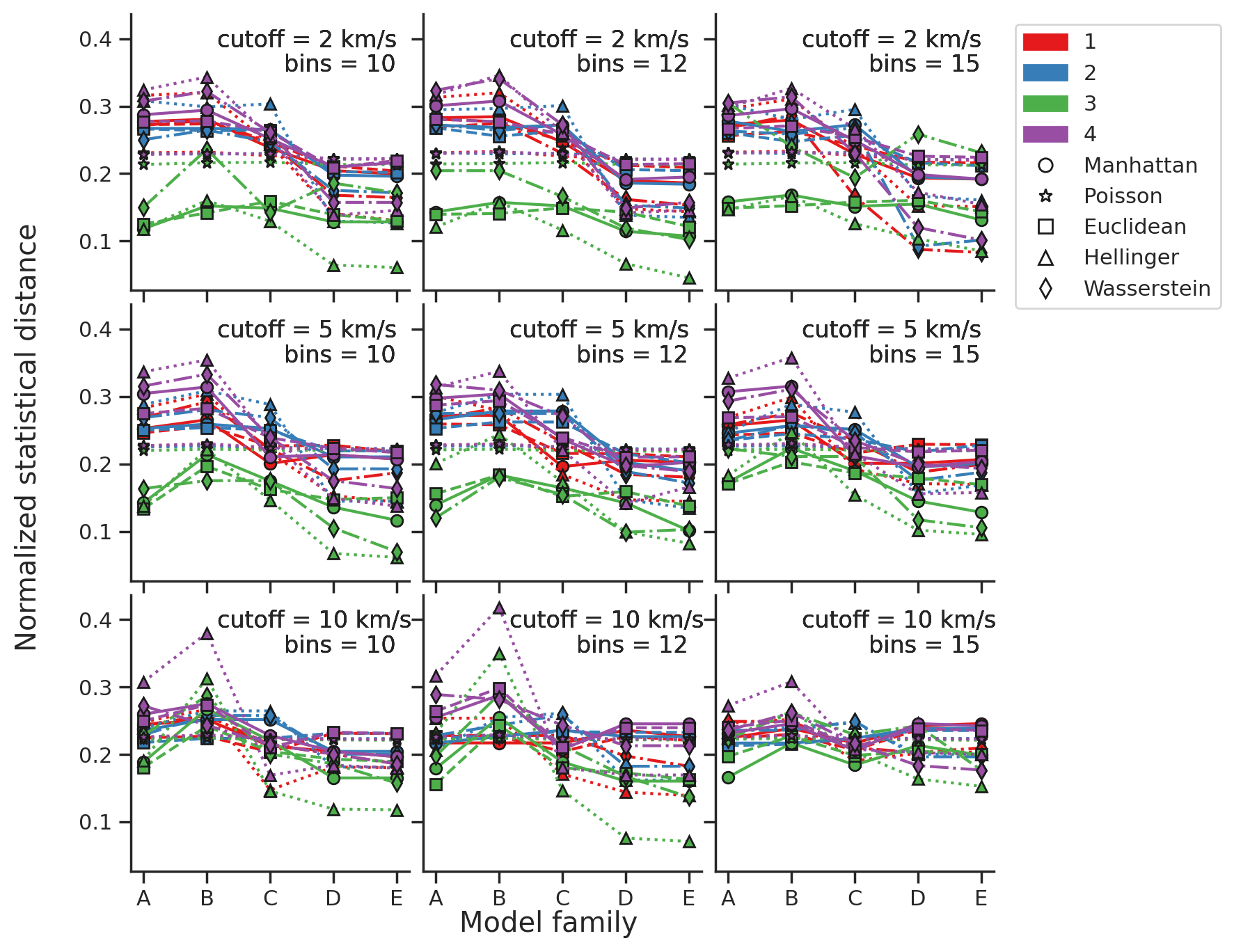}
    \caption{Normalized statistical distances between our COSMIC models and the AGGC \drv distribution, with varying cutoff values for the minimum \drv and number of bins in the compared histograms (histograms shown in Figure~\ref{fig:drvmax_porb_data_all}). The colors indicate the \qcf value, while the markers indicate different statistical measures. Models with \qcf = 3 and models with $\alpha = 5.0$ (D and E) consistently show lower distances to the observational data, which corroborates our finding that these models are better matched to the available data.}
    \label{fig:stat_distances}
\end{figure*}

We also explore different cutoff choices for low \drv and binning for the histograms. A summary of these choices for all metrics is shown in Figure~\ref{fig:stat_distances}. All metrics, regardless of cutoff and binning choice, appear to slightly prefer models with \qcf = 3 as the most similar to the AGGC \drv distribution. 

%
%

\bibliography{bib.bib}

\end{document}